\documentclass[review]{elsarticle}

\usepackage{lineno,hyperref}

\usepackage{amsmath,bm,geometry}
\modulolinenumbers[5]
\journal{Elsevier}

\begin{document}

\begin{frontmatter}

\title{Simplification of the Flux Function for a Higher-order Gas-kinetic Evolution Model}

\author[mymainaddress]{Guangzhao Zhou\corref{mycorrespondingauthor}}
\cortext[mycorrespondingauthor]{Corresponding author}
\ead{zgz@pku.edu.cn}

\author[mysecondaryaddress]{Kun Xu}
\ead{makxu@ust.hk}

\author[mythirdaddress]{Feng Liu}
\ead{fliu@uci.edu}

\address[mymainaddress]{College of Engineering, Peking University, Beijing 100871, China}
\address[mysecondaryaddress]{Department of Mathematics, Hong Kong University of Science and Technology, Clear Water Bay, Kowloon, Hong Kong}
\address[mythirdaddress]{Department of Mechanical and Aerospace Engineering, University of California, Irvine, CA 92697-3975, United States}

\begin{abstract}
The higher-order gas-kinetic scheme for solving the Navier-Stokes equations has been studied in recent years. In addition to the use of higher-order reconstruction techniques, many terms are used in the Taylor expansion of the equilibrium and non-equilibrium gas distribution functions in the higher-order gas kinetic flux function. Therefore, a large number of coefficients need to be determined in the calculation of the time evolution of the gas distribution function at cell interfaces. As a consequence, the higher-order flux function takes much more computational time than that of a second-order gas-kinetic scheme. This paper aims to simplify the evolution model by two steps. Firstly, the coefficients related to the higher-order spatial and temporal derivatives of a distribution function are redefined to reduce the computational cost. Secondly, based on the physical analysis, some terms can be removed without loss of accuracy. As a result, through the simplifications, the computational efficiency of the higher-order scheme is increased significantly. In addition, a self-adaptive numerical viscosity is designed to minimize the necessary numerical dissipation. Several numerical examples are tested to demonstrate the accuracy and robustness of the current scheme.
\end{abstract}

\begin{keyword}
Higher-order scheme \sep Gas-kinetic scheme \sep Gas evolution model
\end{keyword}

\end{frontmatter}


\section{Introduction}

Most of the classical flow solvers are based on the Euler or Navier-Stokes equations. An exact or approximate Riemann solver is usually adopted for the inviscid flux. The viscous flux is treated separately from the inviscid part. However, the gas-kinetic scheme (GKS) \cite{Xu1998, Xu2001} for the computation of compressible flows proceeds from the microscopic dynamic process. The gas distribution function is introduced to follow the gas evolution from a general initial condition in both space and time. Then all macroscopic flow variables are expressed as moments of the distribution function. Since the non-equilibrium part of the distribution function corresponds to the viscous terms, the calculation of the inviscid and viscous fluxes are performed simultaneously.

With the demand for accurate numerical solution and the continuous increase of computational power, more attention has been devoted to the development of higher-order schemes in recent years. At the current stage, the higher-order accuracy in a higher-order method is mostly associated with higher-order reconstruction techniques (e.g., the WENO reconstruction \cite{Liu1994, Jiang1996}), while the Riemann solver stays the same for the flux function as that in low-order schemes. However, since the Riemann solver is intrinsically one dimensional and cannot reflect the variation of variables in the second and third dimension, fluxes have to be evaluated at several Gaussian points on a cell interface to obtain accurate integration in the tangential direction for 2-D and 3-D problems \cite{Titarev2004}. This will require disproportionally more computational cost in comparison with the one-dimensional cases. Recently, the original second-order gas-kinetic scheme is extended to a higher order by several authors \cite{Li2010,Luo2013a,Liu2014}. In addition to the use of the higher-order reconstruction, the evolution process has also a higher-order property. With the Taylor expansions in both perpendicular and tangential directions of a cell interface, multidimensionality is achieved and no Gaussian points are theoretically needed. Further comparison shows that the higher-order evolution model is important in the construction of higher-order schemes \cite{Luo2013b}.

The previous higher-order GKS schemes have shown good performance for both inviscid and viscous flows. However, since the evolution model is associated with a large number of space and time dependent terms in the calculation of the flux, much more computational time is needed than that of a second-order GKS. As reported in Ref. \cite{Luo2013b},  with the same WENO reconstruction, the finite-volume third-order GKS is $4$ times slower than a finite-difference scheme with a Steger-Warming flux splitting method in the two-dimensional simulations. Therefore it is meaningful and necessary to reduce the computational cost of the current higher-order GKS. This paper will follow the main idea in Ref. \cite{Luo2013a} for the construction of WENO-GKS. But, two simplifications are obtained to construct a more efficient scheme without loss of accuracy. The paper is organized as follows. In Section \ref{section method} the general idea of the numerical method is introduced. Section \ref{section previous} is a brief review of the previous higher-order gas-kinetic evolution model as the baseline model. Section \ref{section modify} shows the details of the modifications on the baseline model. Numerical test cases are presented in Section \ref{section cases}. In Section \ref{section efficiency}, we perform the comparison of accuracy and efficiency of the methods before and after the simplifications. Finally, the conclusion is given in Section \ref{section conclusion}.

\section{Numerical Procedure} \label{section method}
We present a brief introduction to the standard procedure of gas-kinetic schemes. More details can be found in Ref. \cite{Xu1998, Xu2001}.

The BGK equation \cite{BGK}:
\begin{equation} \label{bgk}
f_t + \bm{u} \cdot \nabla f = \frac{g-f}{\tau},
\end{equation}
where $f$ is the gas distribution function, $g$ is the equilibrium distribution that $f$ approaches, $\bm{u} = (u,v)^T$ is the particle velocity, and $\tau$ is defined as the collision time (time between collisions). The equilibrium function, known as the Maxwellian distribution, is
\begin{equation} \label{equilibrium}
g = \rho \left( \frac{\lambda}{\pi} \right)^{\frac{K+2}{2}} \!\! e^{-\lambda \left[ (u-U)^2 + (v-V)^2 + \xi^2 \right]}
\end{equation} 
for two-dimensional flow, where $\rho$ is the density, $U$, $V$ are macroscopic velocities in $x$ and $y$ direction, respectively. $\lambda = m/2kT$, where $m$ is the molecular mass, $k$ is the Boltzmann constant and $T$ is the temperature. $K$ is the internal degrees of freedom which equals to $3$ for diatomic molecules. $\xi$ is the internal variable with $\xi^2 = \xi_1^2 + \xi_2^2 + \cdots + \xi_K^2$.

From Eqs. \eqref{bgk} and \eqref{equilibrium}, it is clear that $f$ is a function of $\bm{x}$, $t$, $\bm{u}$ and $\xi$. The macroscopic variables $\rho$, $U$, $V$ and $T$ appear as coefficients that are local constants. The conservative variables are related to the distribution function by the following equation:
\begin{equation} \label{relation1}
\bm{W} = (\rho, \rho U, \rho V, \rho E)^T = \int g \bm{\psi} d\Xi,
\end{equation}
where $E$ is the total energy density. 

Once the distribution function $f$ is obtained, the flux at a vertical (along the $y$ direction) cell interface can be expressed as
\begin{equation} \label{flux}
\bm{F} = \int u f \bm{\psi} d\Xi,
\end{equation}
where $d\Xi = du dv d\xi$, $d\xi = d\xi_1 d\xi_2 \cdots d\xi_K$, and $\bm{\psi}$ is the vector of moments:
\begin{equation}
\bm{\psi} = \left( \psi_1, \psi_2, \psi_3, \psi_4 \right)^T = \left( 1, u, v, \textstyle\frac{u^2 + v^2 + \xi^2}{2} \right)^T.
\end{equation}
In addition, we have
\begin{equation} \label{relation2}
\int \left( g-f \right) \bm{\psi} d\Xi \equiv \bm{0}.
\end{equation}
This is due to the conservation property of $\bm{W}$. It is valid for any $\bm{x}$ and $t$.

For a rectangular cell $[x_{i-1/2},x_{i+1/2}] \times [y_{j-1/2}, y_{j+1/2}]$ with dimensions of $\Delta x_i = x_{i+1/2} - x_{i-1/2}$ and $\Delta y_j = y_{j+1/2} - y_{j-1/2}$, the cell-averaged conservative variable $\bm{W}_{ij}$ is updated from the time $t_n$ to $t_{n+1}$ as follows:
\begin{equation} \label{update}
\begin{aligned}
\bm{W}_{ij}^{n+1} = \bm{W}_{ij}^{n} & - \frac{1}{\Delta x_i \Delta y_j} \int_{t_n}^{t_{n+1}} \int_{-\frac{1}{2} \Delta y_j}^{\frac{1}{2} \Delta y_j} \left[ \bm{F}_{i+1/2	}(t,y)  - \bm{F}_{i-1/2} (t,y) \right] dy dt  \\
&- \frac{1}{\Delta x_i \Delta y_j} \int_{t_n}^{t_{n+1}} \int_{-\frac{1}{2} \Delta x_i}^{\frac{1}{2} \Delta x_i} \left[ \bm{F}_{j+1/2}(t,x) - \bm{F}_{j-1/2} (t,x) \right] dx dt. \\
\end{aligned}
\end{equation}
The above equation is exact. In a conventional high-order finite-volume scheme, the surface and time integration on the right-hand-side is replaced by Gaussian quadrature and multi-step Runge-Kutta schemes of appropriate order, respectively. In the high-order GKS scheme, approximate functions of $\bm{F}$ are obtained by expanding $f$ in $x$, $y$ and $t$, and the surface and time integration is done analytically.

To get $f$, we use the analytical integral solution to Eq. \eqref{bgk}:
\begin{equation} \label{solution}
f(\bm{x},t,\bm{u},\xi) = \frac{1}{\tau} \int_0^t g\left( \bm{x}^\prime, t^\prime,  \bm{u}, \xi \right) e^{-(t-t^\prime) / \tau} dt^\prime + e^{-t/\tau} f_0 \left( \bm{x} - \bm{u} t, \bm{u}, \xi \right),
\end{equation}
where $\bm{x}^\prime = \bm{x} - \bm{u} (t-t^\prime)$ is the particle trajectory. Therefore $f$ depends on the equilibrium distribution function $g$  and the initial distribution function $f_0$. Then the problem is how to model these two functions.

Now we summarize the whole procedure in a time step:
\begin{enumerate}[(1)]
\item Reconstruction. Based on the current values of the averaged conservative variables, construct the values and their spatial derivatives at the midpoint of the cell interface on both sides. This could be done by various techniques. For this paper, the WENO method is applied.
\item Modelling of evolution. Model $f_0$ and $g$ from the reconstructed conservative variables. Then get $f$ via Eq. \eqref{solution}. This paper aims to make this part more efficient.
\item Flux integration. Get the flux at each cell interface according to Eq. \eqref{flux}.
\item Update of conservative variables. This is done following Eq. \eqref{update}.
\end{enumerate}

The reconstruction part can be found in Ref. \cite{Luo2013a}. A standard fifth-order WENO method (WENO-JS) \cite{Jiang1996} is applied in the direction perpendicular to the cell interface to determine the line-averaged values of the variables on both sides of it. Following the suggestion in Ref. \cite{Shu1997}, the characteristic variables are used instead of conservative variables. After that, a third-order interpolation involving the nearby line-averaged values is performed in the tangential direction, to obtain a more accurate value at the midpoint of the interface. Based on the reconstructed variables, the first and second-order derivatives in both $x$ and $y$ directions can be obtained from the reconstructed values and the cell-averaged values, which will be used in the evolution process.

\section{The Baseline Gas-kinetic Evolution Model} \label{section previous}
In this section we introduce the model proposed in Ref. \cite{Luo2013a}, as a baseline for further modifications. For simplicity, we assume that the cell interface is along the $y$ axis, and the midpoint of the interface is located at $y=0$.

Let $g$ denote the Maxwellian distribution function at the point $(x,y,t) = (0,0,0)$. Then $\tilde{g}$, the equilibrium distribution in the neighbourhood can be expressed via the Taylor expansion. To the second order in space and time, it is written as:
\begin{equation} \label{Taylor}
\tilde{g}(\bm{x},t,\bm{u},\xi) = g + g_x x + g_y y + g_t t + \frac{1}{2}g_{xx} x^2 + \frac{1}{2}g_{yy} y^2 + \frac{1}{2}g_{tt} t^2 + g_{xy} xy + g_{xt} xt + g_{yt} yt.
\end{equation}
Introducing the coefficients below:
\begin{equation} \label{coeff}
\begin{aligned}
& a_1 = g_x/g, \quad a_2 = g_y/g, \quad A = g_t/g,  \\
& d_{11} = \frac{\partial a_1}{\partial x},\quad d_{12} = \frac{\partial a_1}{\partial y} = \frac{\partial a_2}{\partial x}, \quad d_{22} = \frac{\partial a_2}{\partial y}, \\
& b_1 = \frac{\partial a_1}{\partial t}  = \frac{\partial A}{\partial x}, \quad  b_2 = \frac{\partial a_2}{\partial t}  = \frac{\partial A}{\partial y}, \quad B = \frac{\partial A}{\partial t},
\end{aligned}
\end{equation}
Eq. \eqref{Taylor} becomes
\begin{equation} \label{g-model}
\begin{aligned}
\tilde{g}(\bm{x},t,\bm{u},\xi) = & g + g a_1 x + g a_2 y + g A t + \frac{1}{2} g \left( a_1^2 + d_{11} \right) x^2 + \frac{1}{2} g \left( a_2^2 + d_{22} \right) y^2 + \frac{1}{2} g \left( A^2 + B \right) t^2 \\
& + g \left( a_1 a_2 + d_{12} \right) xy + g \left( A a_1 + b_1 \right) xt +  g \left( A a_2 + b_2 \right) yt.
\end{aligned}
\end{equation}

Now let's consider the non-equilibrium distribution function $f$. According to the Chapman-Enskog expansion, to the order of  the Navier-Stokes equations , $f$ and $g$ have the following relation \cite{Ohwada2004}:
\begin{equation}
f = g - \tau Dg =  g - \tau \left( g_t + ug_x + vg_y \right).
\end{equation}
Similarly, by applying the Taylor expansion for each term and neglecting high-order derivatives of $g$, we get
\begin{equation}
\begin{aligned}
\tilde{f} (\bm{x},t,\bm{u},\xi) =   & g + g_x x + g_y y + g_t t + \frac{1}{2} g_{xx} x^2 + \frac{1}{2} g_{yy} y^2 + \frac{1}{2} g_{tt}t^2 + g_{xy} xy + g_{xt} xt + g_{yt} yt\\
 & - \tau ( g_t + g_{xt} x  + g_{yt} y + g_{tt}t ) - \tau u ( g_x + g_{xx} x + g_{xy} y + g_{xt} t ) - \tau v ( g_y + g_{xy} x + g_{yy} y + g_{yt} t  ).
\end{aligned}
\end{equation}
With the coefficients defined in Eq. \eqref{coeff}, the expression of the non-equilibrium distribution valid in the neighbour of $(x,y,t) = (0,0,0)$ is:
\begin{equation} \label{f-model}
\begin{aligned}
\tilde{f} (\bm{x},t,\bm{u},\xi) =   & g + g a_1 x + g a_2 y + g A t + \frac{1}{2} g \left( a_1^2 + d_{11} \right) x^2 + \frac{1}{2} g \left( a_2^2 + d_{22} \right) y^2 + \frac{1}{2} g \left( A^2 + B \right) t^2 \\
& + g \left( a_1 a_2 + d_{12} \right) xy + g \left( A a_1 + b_1 \right) xt +  g \left( A a_2 + b_2 \right) yt\\
 & - \tau \left[ gA + g \left( Aa_1 + b_1 \right) x  + g \left( Aa_2 + b_2 \right) y + g\left( A^2+B \right) t  \right] \\
  & - \tau u \left[ ga_1 + g \left( a_1^2 + d_{11} \right) x + g \left( a_1 a_2 + d_{12} \right) y + g \left( A a_1 + b_1 \right) t  \right] \\
 & - \tau v \left[ g a_2 + g \left( a_1 a_2 + d_{12} \right) x + g \left( a_2^2 + d_{22} \right) y + g \left( A a_2 + b_2 \right) t  \right].
\end{aligned}
\end{equation}
Note that for an arbitrarily given equilibrium state $g$, there exist $\tilde{g}$ and $\tilde{f}$ corresponding to $g$. Then we have the form $\tilde{g} = \tilde{g}(g,\bm{x},t,\bm{u})$, $\tilde{f} = \tilde{f}(g,\bm{x},t,\bm{u})$. 
Now model the unknown functions in the solution \eqref{solution} as the following.

The initial state at the cell interface should be discontinuous:
\begin{equation}
f_0 \left( \bm{x}, \bm{u},\xi \right) = \left\{
\begin{aligned}
& f_0^l  \left( \bm{x}, \bm{u},\xi \right) = \tilde{f}^l  \left( g_0^l, \bm{x}, 0, \bm{u}\right) , \quad x \leq 0, \\
& f_0^r  \left( \bm{x}, \bm{u},\xi \right) = \tilde{f}^r  \left( g_0^r, \bm{x}, 0, \bm{u} \right) , \quad x > 0,
\end{aligned}\right.
\end{equation}
where $g_0^l$ and $g_0^r$ correspond to the reconstructed conservative variables at the left and right side of the cell interface, respectively. i.e.,
\begin{equation}
\bm{W}^l = \int g_0^l \bm{\psi} d\Xi, \quad \bm{W}^r = \int g_0^r \bm{\psi} d\Xi.
\end{equation} 

The equilibrium distribution function in the integral solution \eqref{solution} is replaced by
\begin{equation}
g \left( \bm{x}, t, \bm{u},\xi \right)  = \tilde{g} \left( g^e, \bm{x}, t, \bm{u}\right),
\end{equation}
where $g^e$ is formed from the colliding particles from both sides of the interface. According to the relation \eqref{relation1}, it is obtained by:
\begin{equation}
\int g^e \bm{\psi}  d\Xi = \bm{W}^e = \int_{u\geq 0} g_0^l \bm{\psi}  d\Xi+ \int_{u < 0} g_0^r \bm{\psi}  d\Xi.
\end{equation}

By replacing $f_0^{l,r}$ and $g$ with $\tilde{f}^{l,r}$ and $\tilde{g}$, we get an approximation to the exact distribution function $f$ (Eq. \eqref{solution}) . The accuracy of the approximation is related to the order of Taylor expansion in $\tilde{f}^{l,r}$ and $\tilde{g}$. A first-order expansion is enough in the second-order gas-kinetic scheme \cite{Xu2001}. For the current higher-order scheme, we employ the second-order expansion to more accurately reflect the variation of the distribution functions with $x$, $y$ and $t$.

The coefficients $a_1,a_2,A,\cdots$ in the distribution functions \eqref{g-model} and \eqref{f-model} are determined by conservative variables. Each coefficient can be written as $\Lambda = \Lambda_1 \psi_1 +  \Lambda_2 \psi_2 + \Lambda_3\psi_3 + \Lambda_4\psi_4$. Define the moment of a variable as:
\begin{equation}
\left\langle \cdots \right\rangle = \int g(\cdots) \bm{\psi} d\Xi,
\end{equation}
with the help of the relation \eqref{relation2}, they can be derived as follows:
\begin{equation} \label{decide_coef}
\begin{aligned}
& \left\langle a_1 \right\rangle = \bm{W}_x \rightarrow a_1, \quad \left\langle a_2 \right\rangle = \bm{W}_y \rightarrow a_2, \quad \left\langle a_1 u + a_2 v + A \right\rangle = \bm{0} \rightarrow A, \\
& \left\langle a_1^2 + d_{11} \right\rangle = \bm{W}_{xx} \rightarrow d_{11}, \quad  \left\langle a_2^2 + d_{22} \right\rangle = \bm{W}_{yy} \rightarrow d_{22}, \quad \left\langle a_1 a_2 + d_{12} \right\rangle = \bm{W}_{xy} \rightarrow d_{12}, \\
& \left\langle \left( a_1^2 + d_{11} \right) u + \left( a_1 a_2 + d_{12} \right) v + A a_1 + b_1 \right\rangle = \bm{0} \rightarrow b_1, \\
& \left\langle \left( a_1 a_2 + d_{12} \right) u + \left( a_2^2 + d_{22} \right) v + A a_2 + b_2 \right\rangle = \bm{0} \rightarrow b_2, \\
& \left\langle \left( A a_1 + b_1 \right) u + \left( A a_2 + b_2 \right) v + A^2 + B \right\rangle = \bm{0} \rightarrow B.
\end{aligned}
\end{equation}
All moments can be calculated explicitly. Details can be found in Ref. \cite{Xu2001}.

With all the above preparations, the final expression of $f$ at $x=0$ is written as (see Ref. \cite{Luo2013a}):
\begin{equation} \label{result1}
\begin{aligned}
f(0,y,t,\bm{u},\xi) = &  \frac{1}{\tau} \int_0^t g\left( -u \left( t - t^\prime \right), y-v \left( t - t^\prime \right), t^\prime,  \bm{u}, \xi \right) e^{-(t-t^\prime) / \tau} dt^\prime \\
& + e^{-t/\tau} f_0 \left( - ut, y - vt, \bm{u}, \xi \right),
\end{aligned}
\end{equation}
where
\begin{equation} \label{result2}
\begin{aligned}
 & \frac{1}{\tau} \int_0^t g\left( -u \left( t - t^\prime \right), y-v \left( t - t^\prime \right), t^\prime,  \bm{u}, \xi \right) e^{-(t-t^\prime) / \tau} dt^\prime \\
 = &  C_1 g^e +  C_2 g^e a^e_1 u + C_1 g^e a^e_2 y + C_2 g^e a^e_2 v + C_3 g^e A^e + \frac{1}{2} C_4 g^e \left( (a^e_1)^2 + d^e_{11} \right) u^2 + \frac{1}{2} C_1 g^e \left( (a^e_2)^2 + d_{22} \right) y^2  \\
& + C_2 g^e \left( (a^e_2)^2 + d^e_{22} \right) vy + \frac{1}{2} C_4 g^e \left( (a^e_2)^2 + d^e_{22} \right) v^2 + C_2 g^e \left( a^e_1 a^e_2 + d^e_{12} \right) uy +  C_4 g^e \left( a^e_1 a^e_2 + d^e_{12} \right) uv \\
& + \frac{1}{2} C_5 g^e \left( (A^e)^2 + B^e \right) + C_6 g^e \left( A^e a^e_1 + b^e_1 \right) u + C_3 g^e \left( A^e a^e_2 + b^e_2 \right) y + C_6 g^e \left( A^e a^e_2 + b^e_2 \right) v,
\end{aligned}
\end{equation}
and
\begin{equation} \label{result3}
 e^{-t/\tau} f_0 \left( - ut, y - vt, \bm{u}, \xi \right)  = \left\{ 
\begin{aligned}
&  e^{-t/\tau} f_0^l \left( - ut, y - vt, \bm{u}, \xi \right), \quad u\geq 0,  \\
& e^{-t/\tau} f_0^r \left( - ut, y - vt, \bm{u}, \xi \right), \quad u<0, 
\end{aligned}  \right.
\end{equation}
where
\begin{equation} \label{result4}
\begin{aligned}
& e^{-t/\tau} f_0^{l,r} \left( - ut, y - vt, \bm{u}, \xi \right) \\
= & C_7 g_0^{l,r} \left[ 1 - \tau \left( a^{l,r}_1 u + a_2^{l,r} v + A^{l,r} \right) \right] \\
& + C_8 g_0^{l,r} \left[ a_1^{l,r} u - \tau \left( \left( \left( a_1^{l,r} \right)^2 + d_{11}^{l,r} \right) u^2 + \left( a_1^{l,r} a_2^{l,r} + d_{12}^{l,r} \right) uv + \left( A^{l,r} a_1^{l,r} + b_1^{l,r} \right) u \right) \right]  \\
& + C_7 g_0^{l,r}  \left[ a_2^{l,r} - \tau \left( \left( a_1^{l,r} a_2^{l,r} + d_{12}^{l,r} \right) u + \left( \left( a_2^{l,r} \right)^2 +d_{22}^{l,r} \right) v + A^{l,r} a_2^{l,r} + b_2^{l,r} \right) \right] y \\
& + C_8 g_0^{l,r} \left[ a_2^{l,r} v - \tau \left( \left(  a_1^{l,r} a_2^{l,r} + d_{12}^{l,r} \right) uv + \left( \left(a_2^{l,r} \right)^2 + d_{22}^{l,r} \right) v^2 + \left( A^{l,r} a_2^{l,r} + b_2^{l,r} \right) v \right) \right]  \\
& + \frac{1}{2} C_9 g_0^{l,r} \left( \left( a_1^{l,r} \right)^2 + d_{11}^{l,r} \right) u^2 + \frac{1}{2} C_7 g_0^{l,r} \left( \left( a_2^{l,r} \right)^2 + d_{22}^{l,r} \right) y^2 + C_8 g_0^{l,r} \left( \left( a_2^{l,r} \right)^2 + d_{22}^{l,r} \right) vy \\
& + \frac{1}{2} C_9 g_0^{l,r} \left( \left( a_2^{l,r} \right)^2 + d_{22}^{l,r} \right) v^2  + C_8 g_0^{l,r} \left( a_1^{l,r} a_2^{l,r} + d_{12}^{l,r} \right) uy +  C_9 g_0^{l,r} \left( a_1^{l,r} a_2^{l,r} + d_{12}^{l,r} \right) uv. 
\end{aligned}
\end{equation}
The coefficients are
\begin{equation} \label{result5}
\begin{aligned}
& C_1 = 1 - e^{-t/\tau_n}, \quad C_2 = \left( t+ \tau \right) e^{-t/\tau_n} - \tau, \quad C_3 = t - \tau + \tau e^{-t/\tau_n}, \\
& C_4 = \left( -t^2 - 2\tau t \right) e^{-t/\tau_n},\quad  C_5 = t^2 - 2 \tau t,\quad C_6 = -\tau t \left( 1 +  e^{-t/\tau_n} \right), \\
& C_7 = e^{-t/\tau_n}, \quad C_8 = -t e^{-t/\tau_n}, \quad C_9 = t^2 e^{-t/\tau_n},
\end{aligned}
\end{equation}
where $\tau_n$ is a numerical collision time \cite{Luo2013a}.

\section{Modifications and Discussions on the Baseline Evolution Model} \label{section modify}

\subsection{Simplification 1}
The first simplification is to rearrange the coefficients in formulae \eqref{coeff}.  Instead of using \eqref{coeff}, a new set of coefficients is introduced:
\begin{equation} \label{coeff_new}
\begin{aligned}
& a_{x} = g_x/g, \quad a_{y} = g_y/g, \quad a_{t} = g_t/g,  \\
& a_{xx} = g_{xx}/g, \quad a_{yy} = g_{yy}/g,  \quad a_{xy} = g_{xy}/g,  \\
& a_{xt} = g_{xt}/g, \quad a_{yt} = g_{yt}/g, \quad a_{tt} = g_{tt}/g.
\end{aligned}
\end{equation}
Then formulae \eqref{decide_coef} are replaced by
\begin{equation} \label{decide_coeff_new}
\begin{aligned}
& \left\langle a_{x} \right\rangle = \bm{W}_x \rightarrow a_{x}, \quad \left\langle a_y \right\rangle = \bm{W}_y \rightarrow a_y, \quad \left\langle a_x u + a_y v + a_t \right\rangle = \bm{0} \rightarrow a_t, \\
& \left\langle a_{xx} \right\rangle = \bm{W}_{xx} \rightarrow a_{xx}, \quad  \left\langle a_{yy} \right\rangle = \bm{W}_{yy} \rightarrow a_{yy}, \quad \left\langle a_{xy} \right\rangle = \bm{W}_{xy} \rightarrow a_{xy}, \\
& \left\langle a_{xx} u +  a_{xy} v + a_{xt} \right\rangle = \bm{0} \rightarrow a_{xt}, \quad \left\langle a_{xy} u +  a_{yy} v + a_{yt} \right\rangle = \bm{0} \rightarrow a_{yt}, \\
& \left\langle a_{xt} u +  a_{yt} v + a_{tt} \right\rangle = \bm{0} \rightarrow a_{tt}.
\end{aligned}
\end{equation}

The modification has the following properties:
\begin{enumerate}[(1)]
\item The new definition does not change the total number of coefficients. The degree of freedom remains the same. Since the coefficients are used to get a first-order approximation in terms of $\bm{\psi}$, this modification will not cause loss of accuracy.
\item Comparing Eq. \eqref{decide_coeff_new} with Eq. \eqref{decide_coef}, it is clear that the computational cost of determining the coefficients are reduced a lot.
\item With the new coefficients, the solution $f$ will be much simplified. All combined coefficients in Eqs. \eqref{result2} and \eqref{result4} are replaced by a single coefficient. They are:
\begin{equation}
\begin{aligned}
& a_1^2 + d_{11} \rightarrow a_{xx}, \quad a_2^2 + d_{22} \rightarrow a_{yy}, \quad a_1 a_2 + d_{12} \rightarrow a_{xy},
\\
& A a_1 + b_1 \rightarrow a_{xt}, \quad A a_2 + b_2 \rightarrow a_{yt}, \quad A^2 + B \rightarrow a_{tt} .
\end{aligned}
\end{equation}
Notice that each coefficient is a summation of 4 terms, e.g.,
\begin{equation} \label{simple_expansion}
 a_{tt} = a_{tt}^{(1)} + a_{tt}^{(2)} u + a_{tt}^{(3)} v + a_{tt}^{(4)} \frac{u^2 + v^2 + \xi^2}{2},
\end{equation}
as a comparison, $A^2 + B$ is expanded as
\begin{equation} \label{complex_expansion}
  \left( A^{(1)} + A^{(2)} u + A^{(3)} v + A^{(4)} \textstyle\frac{u^2 + v^2 + \xi^2}{2} \right) ^2 +  \left( B^{(1)} + B^{(2)} u + B^{(3)} v + B^{(4)} \textstyle\frac{u^2 + v^2 + \xi^2}{2} \right),
\end{equation}
which contains many more terms than $4$. So the time saved is considerable.
\item The original definition of coefficients will lead to high-order terms of $\bm{\psi}$. In the expansion of $A^2 + B$, the highest power on $u$, $v$ and $\xi$ is $4$. While in the expansion of $a_{tt}$, the highest power is $2$. Without involving very high moments of the distribution function, the simplified scheme tends to be more stable.

\end{enumerate}

\subsection{Simplification 2}
By substituting the coefficients and rearranging the terms, the final solution of the distribution function $f$ at the cell interface (Eqs. \eqref{result1} - \eqref{result5}) can be expressed as following,
\begin{equation} \label{before_simplification}
\begin{aligned}
& f(0,y,t,\bm{u},\xi)  \\
= &  \frac{1}{\tau} \int_0^t g\left( -u \left( t - t^\prime \right), y-v \left( t - t^\prime \right), t^\prime,  \bm{u}, \xi \right) e^{-(t-t^\prime) / \tau} dt^\prime + e^{-t/\tau} f_0 \left( - ut, y - vt, \bm{u}, \xi \right) \\
= & g^e + \frac{1}{2} g^e_{yy} y^2 + g^e_t t + \frac{1}{2} g^e_{tt} t^2 - \tau \left[ \left( g^e_t + u g^e_x + v g^e_y \right) + \left( g^e_{tt} + u g^e_{xt} + v g^e_{yt} \right) t \right] \\
& - e^{-t/\tau} \left[ \begin{aligned}
 g^e & + \frac{1}{2} g^e_{yy} y^2 -  \left( u g^e_x + v g^e_y \right) t + \frac{1}{2} \left( u^2 g^e_{xx} + 2uv g^e_{xy} + v^2 g^e_{yy} \right) t^2 \\
& - \tau \left[ \left( g^e_t + u g^e_x + v g^e_y \right) - \left( u g^e_{xt} + v g^e_{yt} + u^2 g^e_{xx} + 2uv g^e_{xy} + v^2 g^e_{yy} \right) t \right] 
\end{aligned}\right] \\
& + e^{-t/\tau} \left\{ \begin{aligned}
g^l & + \frac{1}{2} g^l_{yy} y^2 -  \left( u g^l_x + v g^l_y \right) t + \frac{1}{2} \left( u^2 g^l_{xx} + 2uv g^l_{xy} + v^2 g^l_{yy} \right) t^2 \\
& - \tau \left[ \left( g^l_t + u g^l_x + v g^l_y \right) - \left( u g^l_{xt} + v g^l_{yt} + u^2 g^l_{xx} + 2uv g^l_{xy} + v^2 g^l_{yy} \right) t \right],\quad u \geq 0, \\
 g^r & + \frac{1}{2} g^r_{yy} y^2 -  \left( u g^r_x + v g^r_y \right) t + \frac{1}{2} \left( u^2 g^r_{xx} + 2uv g^r_{xy} + v^2 g^r_{yy} \right) t^2 \\
& - \tau \left[ \left( g^r_t + u g^r_x + v g^r_y \right) - \left( u g^r_{xt} + v g^r_{yt} + u^2 g^r_{xx} + 2uv g^r_{xy} + v^2 g^r_{yy} \right) t \right], \quad u<0. 
\end{aligned} \right. \\
& +Y.
\end{aligned}
\end{equation}
Here the terms with the factor $y$ is collected into $Y$, i.e., $Y = yH$, where $H$ is independent of $y$. Since the distribution function is to be integrated in the interval $[-\frac{1}{2}\Delta y, \frac{1}{2}\Delta y]$, these terms will vanish. So it is unnecessary to include them. Note that the expression Eq. \eqref{before_simplification} is the original physical one, we should keep in mind that all $e^{-t/\tau}$ is actually $e^{-t/\tau_n}$ in practical computations.

Consider the expression of $f$, the terms without the factor $e^{-t/\tau}$ are exactly a combination of the Chapman-Enskog expansion and the Taylor expansion based on $g^e$ at $(x,y,t) = (0,0,0)$. While the terms with the factor $e^{-t/\tau}$ have the same form for the initial discontinuous (left and right) state and the equilibrium state. But the sign in front of them are different. Therefore $f$ can be written formally as
\begin{equation}
f = \widetilde{g^e} - \tau \widetilde{Dg^e} + e^{-t/\tau} \left( E^{l,r} - E^e\right).
\end{equation}

\begin{enumerate}[(1)]
\item In Euler cases, $\tau = 0$, the exponential part vanishes.
\item In viscous cases, when the mesh for computation is fine enough, the flow field is locally smooth, hence the variables and their gradients at the left/right and the equilibrium states are identical: $E^{l,r} = E^e$. Then the exponential part vanishes.
\item The exponential part makes sense only when there is a discontinuity at the cell interface.
\end{enumerate}
Since for viscous flows an absolute discontinuity does not actually exist in the real world, the flow field will be smooth everywhere if very fine grids are adopted, in which case the exponential part disappears. Then it can be concluded that the dominating part of the distribution function is $ \widetilde{g^e} - \tau \widetilde{Dg^e}$, while $e^{-t/\tau} \left( E^{l,r} - E^e\right)$ plays a role of numerical dissipation, which is related to the relative scale of the computational mesh and the physical structure thickness. In other words, the function of these terms is to suppress oscillations near discontinuities on coarse grids. Hence they can be simplified appropriately.

From another aspect, it is well known that almost all reconstruction techniques including WENO will encounter an order-reduction at discontinuities, and non-physical artificial viscosity is generally needed at such regions. Remember that the term $e^{-t/\tau_n}$ is practically used in place of $e^{-t/\tau}$. There is no a unique theory for the construction of $\tau_n$. These uncertainties make it meaningless to get very accurate values for the terms in the bracket behind $e^{-t/\tau_n}$. Then it is reasonable to just keep the primary terms in $ E^{l,r} - E^e$ for providing necessary numerical dissipation.

Notice that $E^{e,l,r}$ contains both inviscid and viscous terms (which are related to $\tau$). Since the viscosity generally plays the role of smoothing the flow field, the physical viscous effect is not dominant wherever a discontinuity exists. Then it is reasonable to get rid of all terms with the factor $\tau$.  On the other hand, we can neglect all second-order terms associated with small parameters $y$ and $t$. The resultant expression is
 \begin{equation} \label{simplifiedE}
E^{e,l,r} =  g^{e,l,r} -  \left( u g^{e,l,r}_x + v g^{e,l,r}_y \right) t. 
\end{equation} 
Our numerical results show that the first-order terms of $t$ are necessary thus this form could not be further simplified. From various test cases, the differences between the results obtained with the form in Eq. \eqref{simplifiedE} and those obtained with the original form in Eq. \eqref{before_simplification} are very small. And neither of the two forms is superior to the other in terms of accuracy and mesh convergence.

Then we conclude that the final expression of the new modelling of the distribution function is:
\begin{equation} \label{simplified}
\begin{aligned}
 f(0,y,t,\bm{u},\xi)  =& g^e + \frac{1}{2} g^e_{yy} y^2 + g^e_t t + \frac{1}{2} g^e_{tt} t^2 - \tau \left[ \left( g^e_t + u g^e_x + v g^e_y \right) + \left( g^e_{tt} + u g^e_{xt} + v g^e_{yt} \right) t \right] \\
& - e^{-t/\tau} \left[ g^e  -  \left( u g^e_x + v g^e_y \right) t  \right]  + e^{-t/\tau} \left\{ \begin{aligned}
g^l & -  \left( u g^l_x + v g^l_y \right) t ,\quad u \geq 0\\
 g^r & -  \left( u g^r_x + v g^r_y \right) t , \quad u<0
\end{aligned} \right\}  + Y.
\end{aligned}
\end{equation}
Obviously Eq. \eqref{simplified} is much simpler than the original one Eq. \eqref{before_simplification}. In next sections we will prove that the new method has a good performance for different kinds of test cases.

\subsection{The numerical collision time}
For Navier-Stokes solutions, the numerical collision time $\tau_n$ is also modified. In Ref. \cite{Luo2013a} the formula is written as:
\begin{equation}
\tau = \frac{\mu}{p^e}, \quad \tau_n = \tau + \beta \Delta x \sqrt{\lambda^e}   \left| \frac{p^l - p^r}{p^l + p^r} \right|,
\end{equation}
where $\mu$ is the dynamic viscosity at the cell interface and $p^e$ and $\lambda^e$ take values corresponding to the equilibrium state. As is known, $\tau_n$ includes both physical and numerical dissipation. We found that since the scheme itself has the mechanism to suppress oscillations near discontinuities, extra artificial viscosity is needed only when the discontinuity is huge. Then the following formula is designed:
\begin{equation}
\tau_n = \tau + \alpha \Delta t e^{1 - \eta^{-10}}, \quad \eta = \left| \frac{p^l - p^r}{p^l + p^r} \right|.
\end{equation}
where $\alpha$ is a constant. It is equal to $0.3$ in our computations.

For one-dimensional Euler cases, the setting $\tau = \tau_n = 0$ will not cause any problem. For two-dimensional Euler cases, some numerical dissipation is needed. The physical collision time is still $\tau = 0$. For the numerical collision time $\tau_n$, we use the formula in Ref. \cite{Liu2014}, which reads
\begin{equation}
\tau_n = C_1 \Delta t + C_2 \Delta t \eta,
\end{equation} 
where $C_1 = 0.1$ and $C_2 = 1$.

\section{Numerical Results} \label{section cases}
The cases tested are of different categories, including one-dimensional and two-dimensional flows, inviscid and viscous flows, and high-speed and low-speed flows. Most of them are standard test cases for high-order schemes. The baseline method has a third-order accuracy for 2-D flows according to Ref. \cite{Luo2013a}. From various numerical cases, the results of the present method are similar to those in Ref. \cite{Luo2013a, Luo2013b, Pan2016}. For all cases, the current results have a good agreement with the benchmark solutions. 

In the computations, all meshes used are uniform with $\Delta x = \Delta y$. The CFL number is set to be $0.6$ if not specified.

\subsection{1-D test cases}
\begin{figure}[htbp]
\centering
\begin{minipage}[t]{0.5\textwidth}
\flushright
\includegraphics[width=\textwidth]{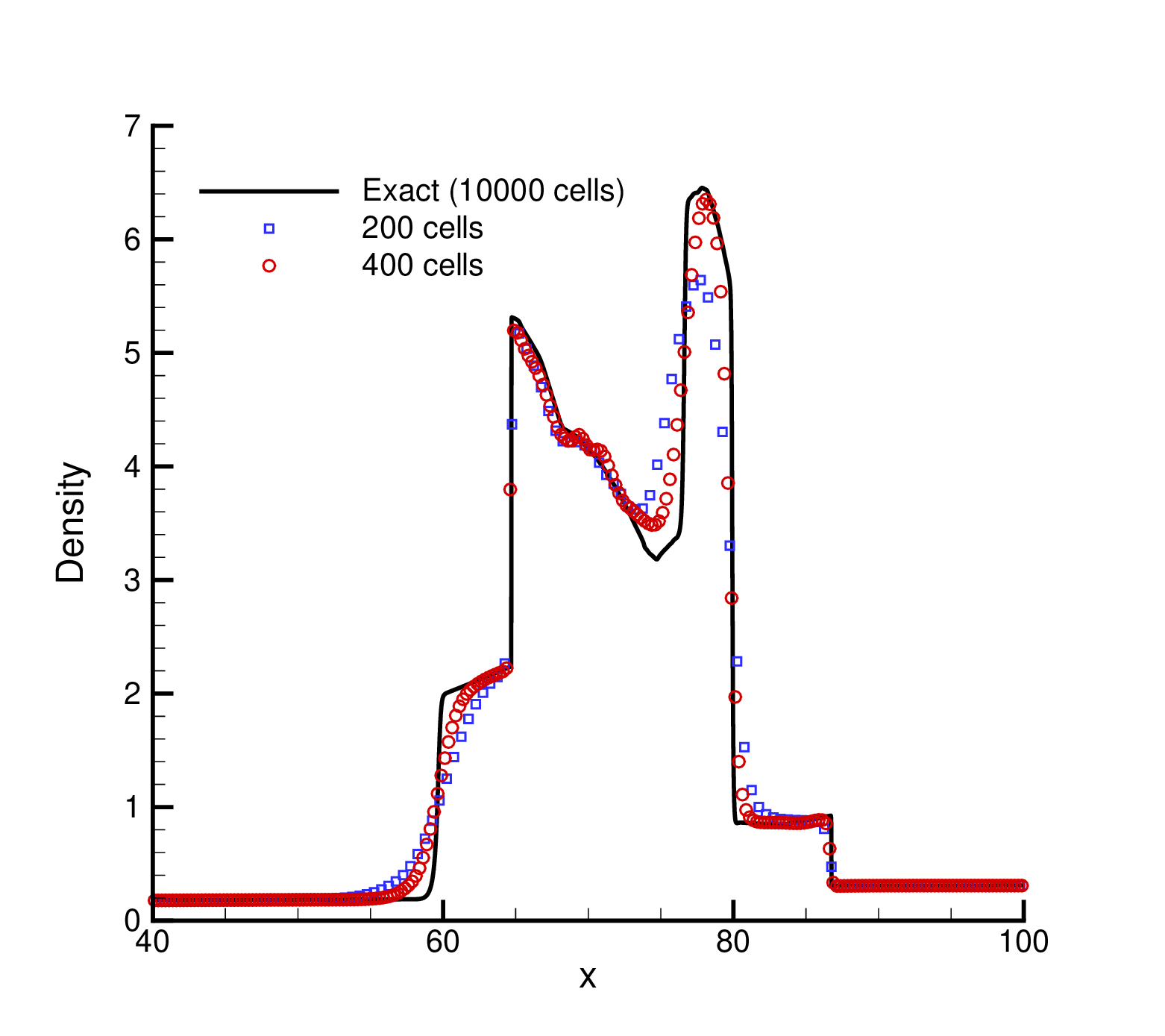}
\end{minipage}%
\begin{minipage}[t]{0.5\textwidth}
\flushleft
\includegraphics[width=\textwidth]{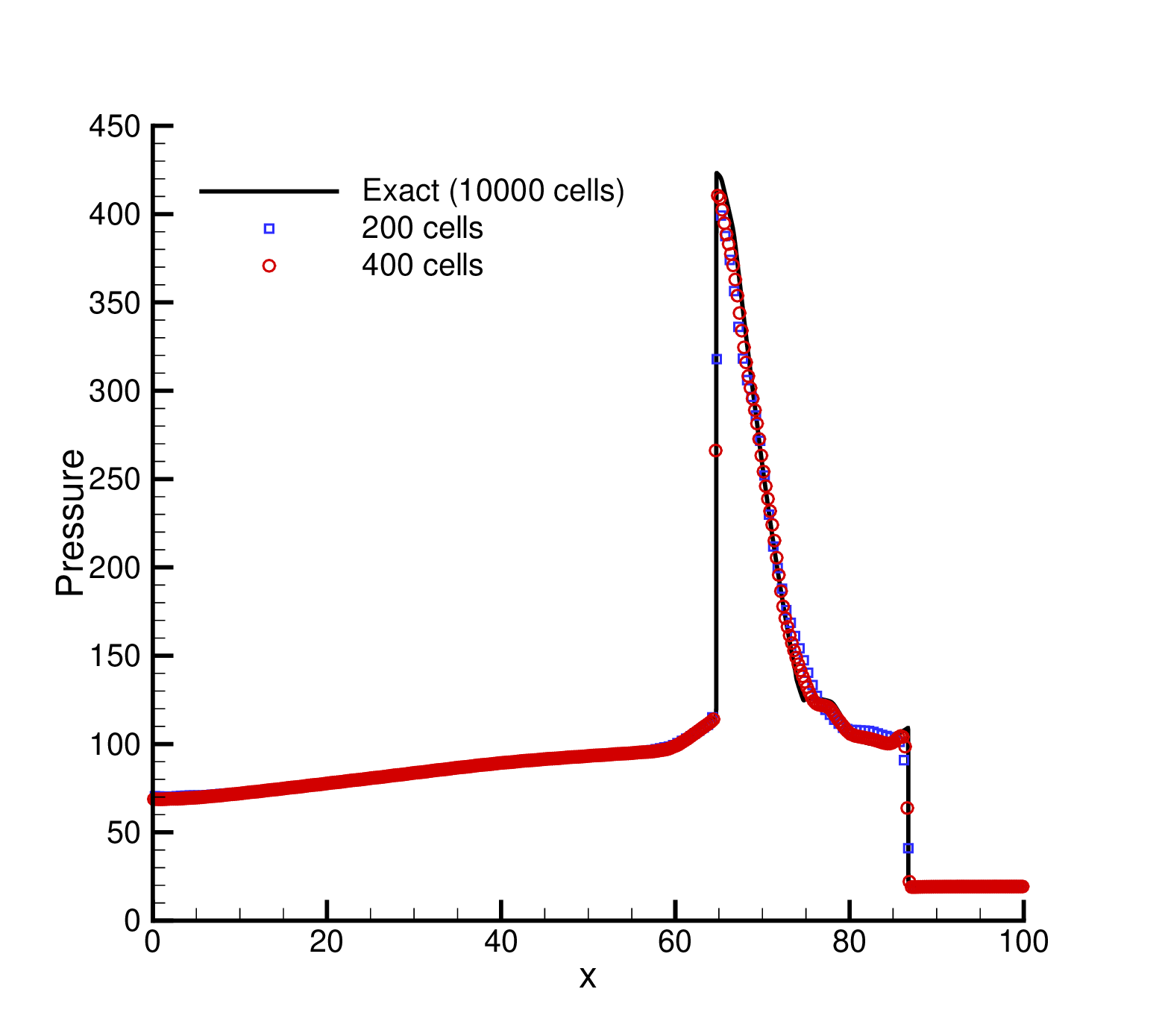}
\end{minipage}
\caption{Density and pressure distributions at $t=3.8$ of the blast wave problem.}\label{blast}
\end{figure}

Two 1-D Riemann problems are computed. The first one is the blast wave problem proposed by Woodward and Colella \cite{Woodward1984}. The initial condition is given by
\begin{equation}
\left( \rho, U, p \right) = \left\{ \begin{aligned}
& \left( 1, 0, 1000 \right), \quad  0 \leq x < 10,  \\
& \left( 1, 0, 0.01 \right), \quad  10 \leq x < 90,  \\
& \left( 1, 0, 100 \right), \quad  90 \leq x \leq 100. 
\end{aligned} \right.
\end{equation}
The density and pressure distributions at $t = 3.8$ are plotted in Fig. \ref{blast}.

\begin{figure}[htbp]
\centering
\begin{minipage}[t]{0.5\textwidth}
\flushright
\includegraphics[width=\textwidth]{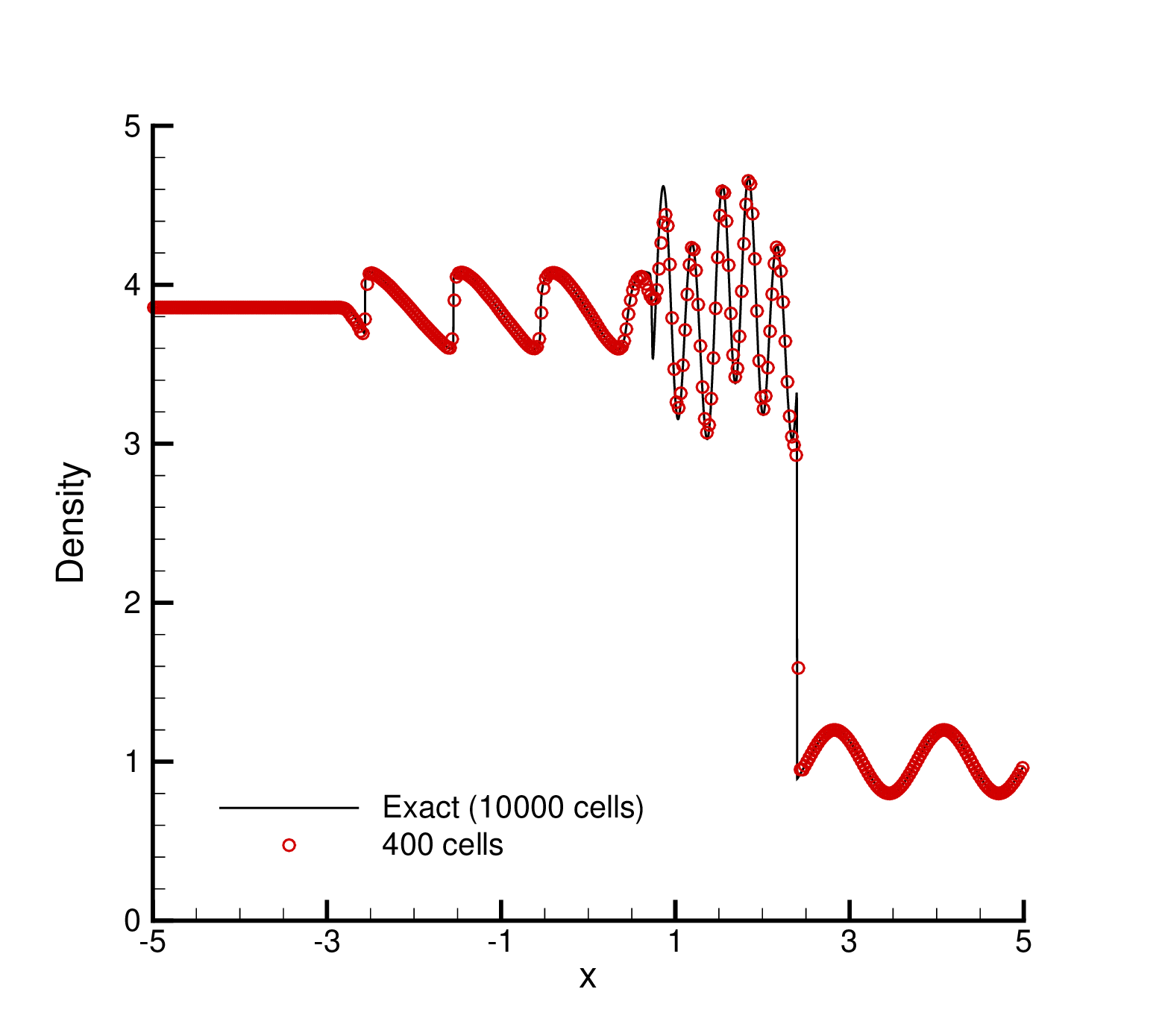}
\end{minipage}%
\begin{minipage}[t]{0.5\textwidth}
\flushleft
\includegraphics[width=\textwidth]{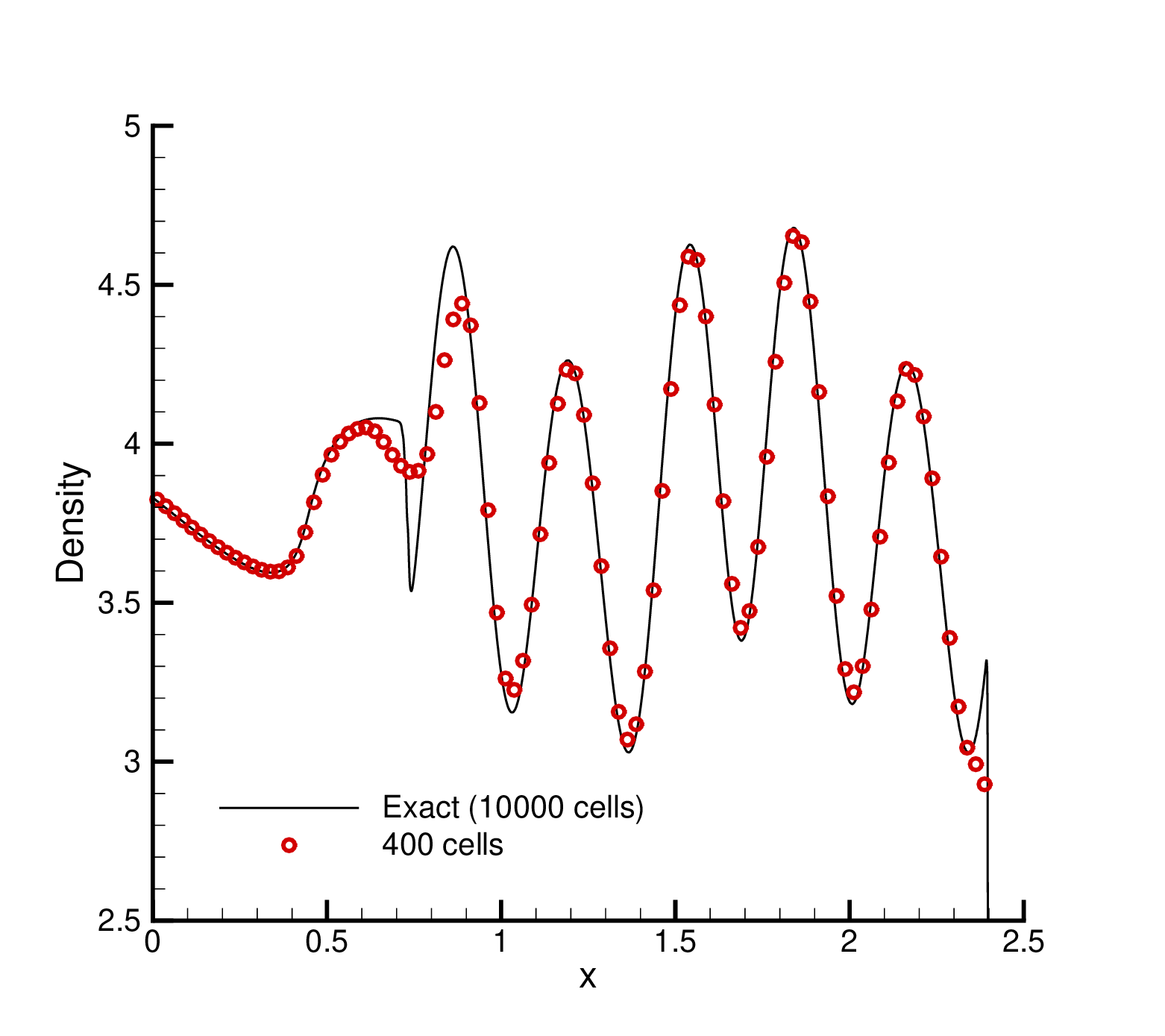}
\end{minipage}
\caption{Density distribution at $t = 1.8$ of the Shu-Osher problem.}\label{shu}
\end{figure}

The second  one is the Shu-Osher problem, which simulates the interaction of a moving shock and a  smooth density fluctuation \cite{Shu1989}.  The initial condition is
\begin{equation}
\left( \rho, U, p \right) = \left\{ \begin{aligned}
& \left( 3.857134, 2.629369, 10.33333 \right), \quad -5 \leq x < -4,  \\
& \left( 1 + 0.2 \sin\left(5 x \right) , 0, 1 \right), \qquad \quad \qquad -4 \leq x \leq 5. 
\end{aligned} \right.
\end{equation}
The density distribution at $t = 1.8$ is shown in Fig. \ref{shu}. The oscillation region is well resolved when $400$ cells are used.

\subsection{Double Mach reflection}
\begin{figure}[htbp]
\centering
\begin{minipage}[t]{\textwidth}
\centering
\includegraphics[width=\textwidth]{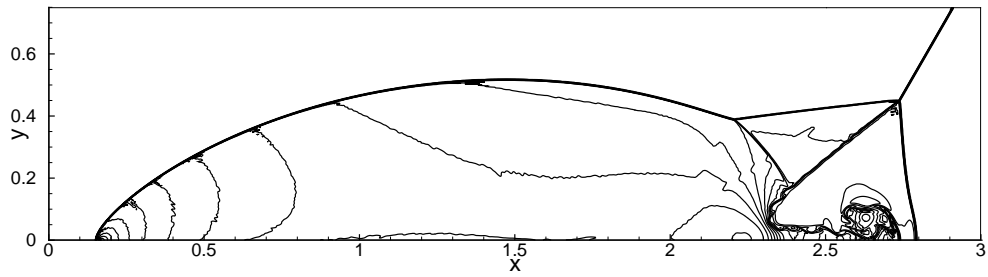}
\end{minipage}
\begin{minipage}[t]{0.5\textwidth}
\flushright
\includegraphics[width=\textwidth]{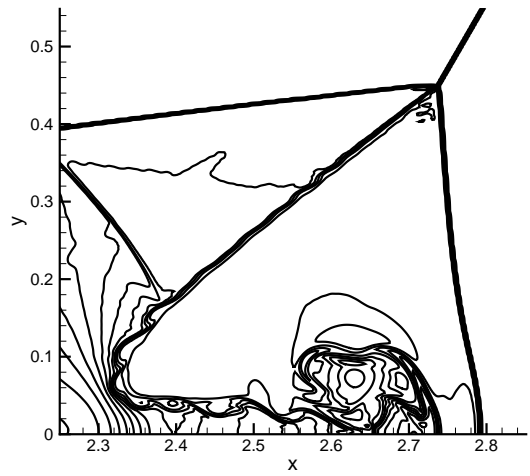}
\end{minipage}
\caption{Double Mach reflection problem. Density distribution at $t = 0.2$ with $\Delta x = \Delta y = 1/480$. $30$ contours equally spaced from $1.5$ to $23$.}\label{dmr}
\end{figure}

The computational domain is $[0,3] \times [0, 0.75]$.  An incident shock wave with $Ma=10$ reflects from the bottom wall starting from $x=1/6$ \cite{Woodward1984}. The adiabatic Euler slip boundary condition is applied at the wall. The exact solutions are used for the post-shock boundaries. The density distribution at $t=0.2$ is shown in Fig. \ref{dmr}.

\subsection{2-D Riemann problem}
\begin{figure}[htbp]
\centering
\begin{minipage}[t]{0.5\textwidth}
\flushright
\includegraphics[width=\textwidth]{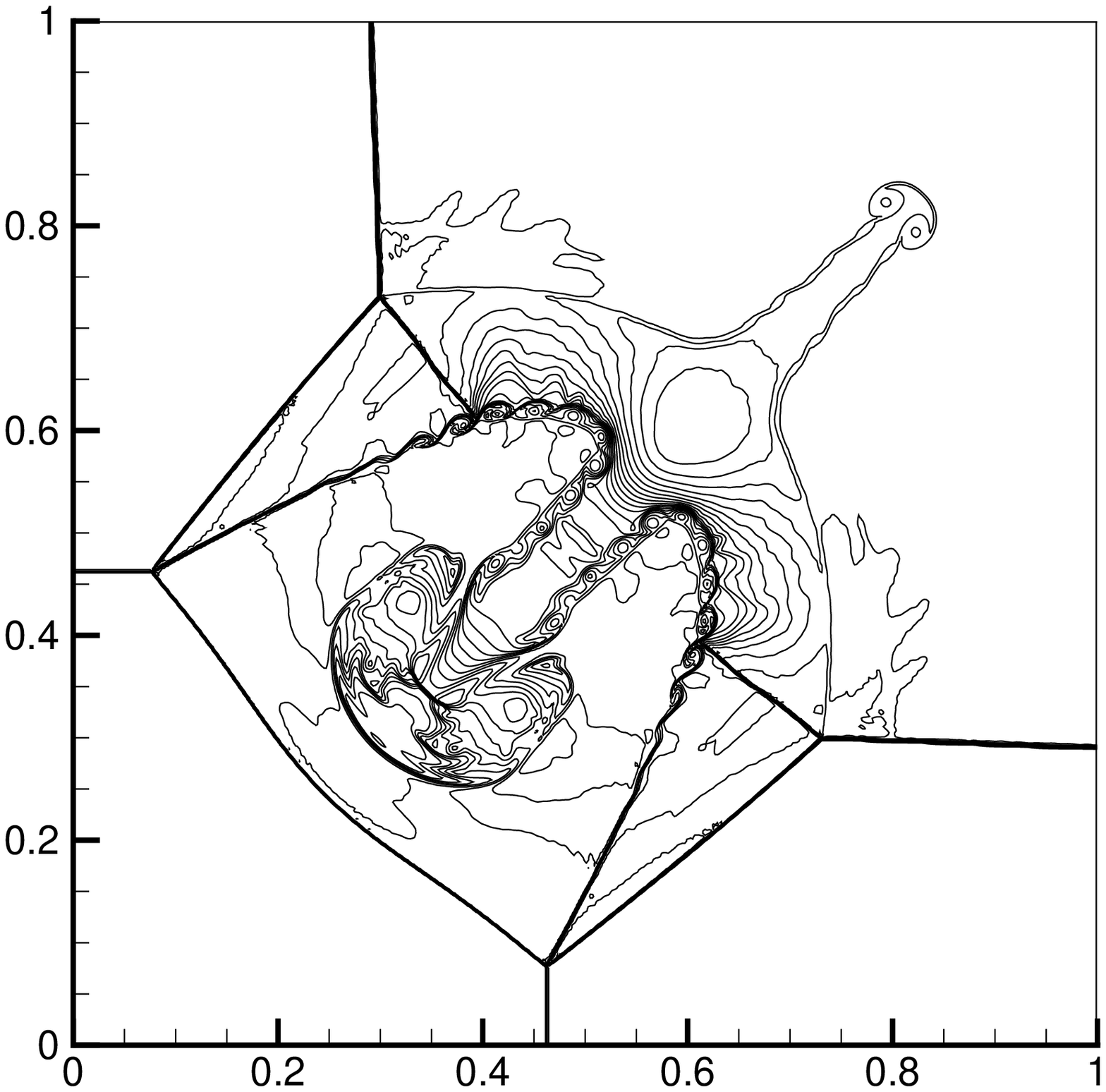}
\end{minipage}%
\begin{minipage}[t]{0.5\textwidth}
\flushleft
\includegraphics[width=\textwidth]{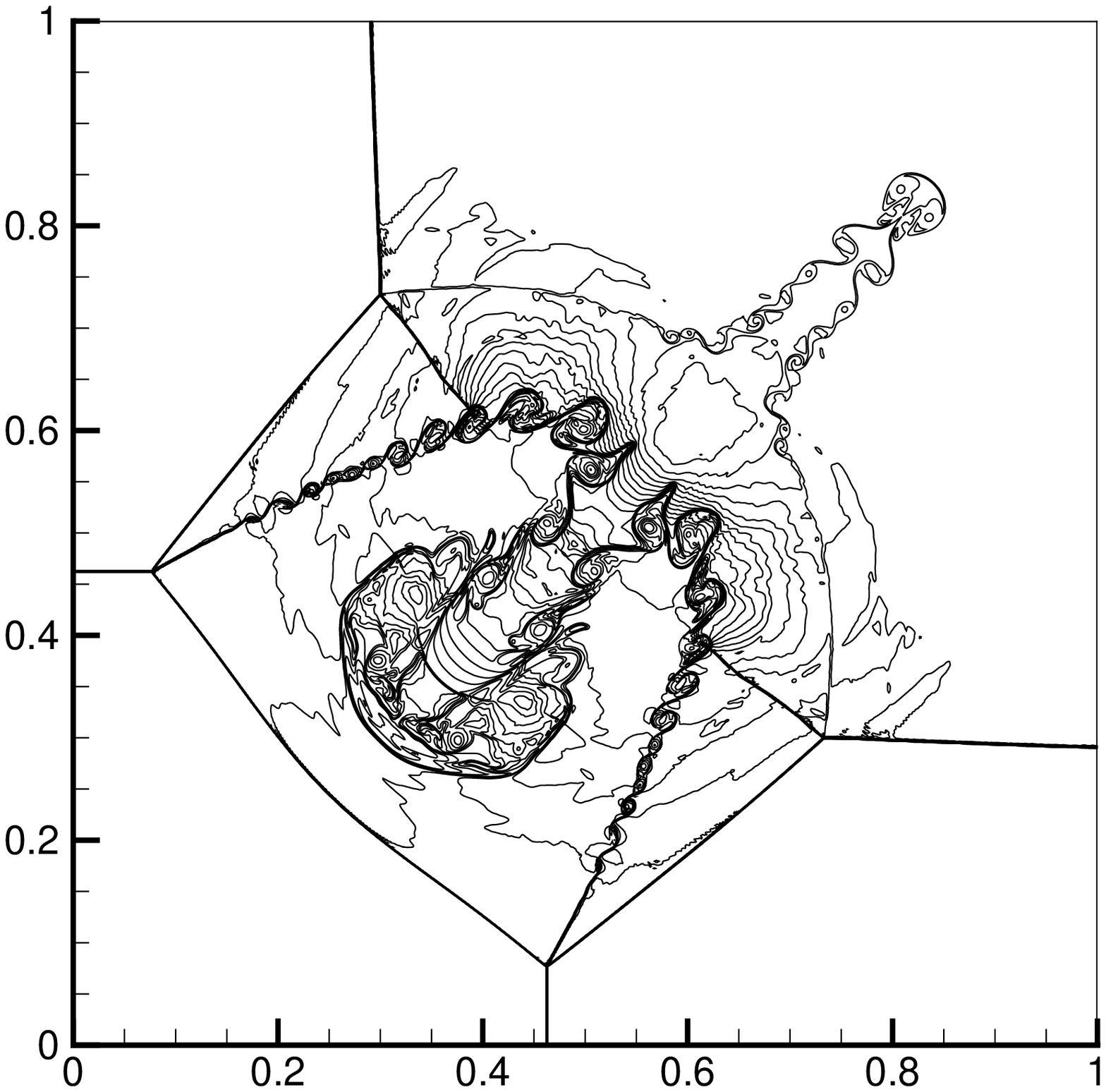}
\end{minipage}
\caption{Density distribution at $t=0.8$ of the 2-D Riemann problem. $25$ contours equally spaced from $0.2$ to $1.7$. $500 \times 500$ mesh (left) and $1000 \times 1000$ mesh (right).}\label{2r_density}
\end{figure}

\begin{figure}[htbp]
\centering
\begin{minipage}[t]{0.5\textwidth}
\flushright
\includegraphics[width=\textwidth]{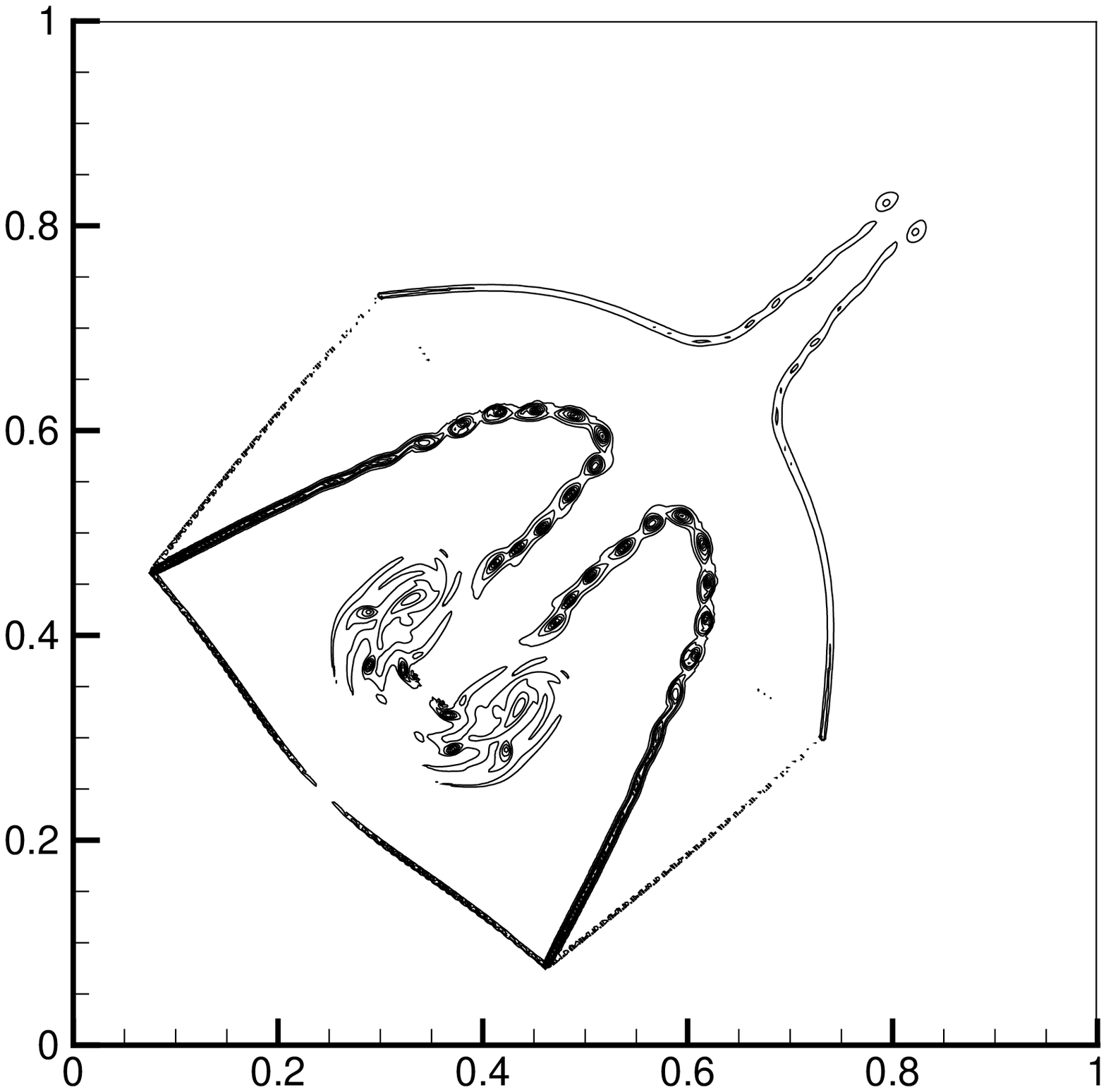}
\end{minipage}%
\begin{minipage}[t]{0.5\textwidth}
\flushleft
\includegraphics[width=\textwidth]{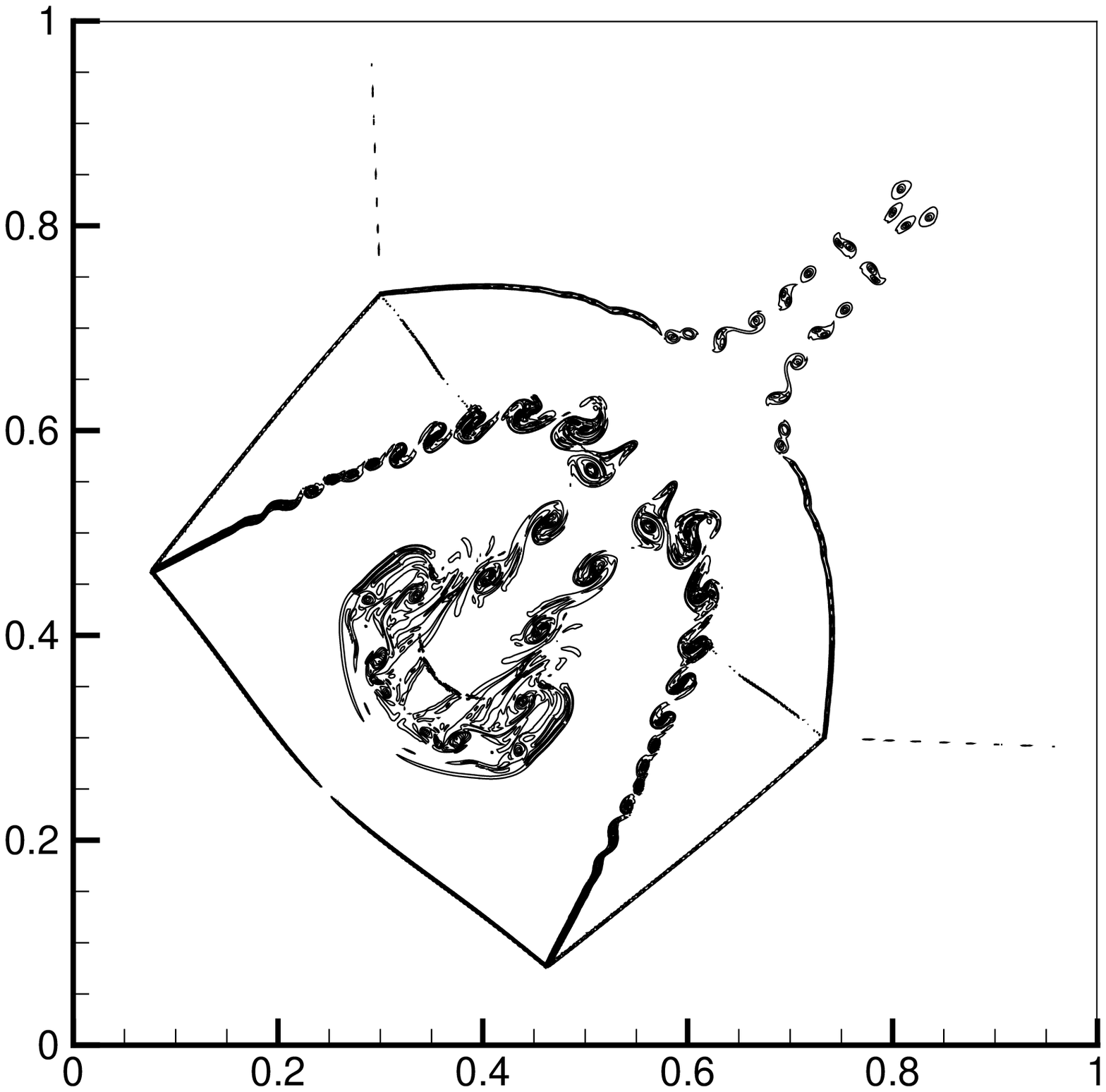}
\end{minipage}
\caption{Vorticity magnitude distribution at $t=0.8$ of the 2-D Riemann problem. $21$ contours equally spaced from $20$ to $600$. $500 \times 500$ mesh (left) and $1000 \times 1000$ mesh (right).}\label{2r_vorticity}
\end{figure}

This case is one of the two-dimensional Riemann problems listed in Ref. \cite{Zhang1990}. The computational domain is $[0,1] \times [0,1]$. We use the initial conditions given in Ref. \cite{benchmark}:
\begin{equation}
\left( \rho, U, V, p \right) = \left\{ \begin{aligned}
& \left( 1.5, 0, 0, 1.5 \right), \quad  x \geq 0.8, y\geq 0.8 ,  \\
& \left( 0.5323, 1.206, 0, 0.3 \right), \quad  x < 0.8, y\geq 0.8 ,  \\
& \left( 0.138, 1.206, 1.206, 0.029 \right), \quad  x < 0.8, y < 0.8, \\ 
& \left( 0.5323, 0, 1.206, 0.3 \right), \quad  x \geq 0.8, y < 0.8.
\end{aligned} \right.
\end{equation}
At $t>0$, four shock waves form at the interfaces of different states. They interact at the junction point and a system of complex structures appear. Two different meshes are used, $500 \times 500$ and $1000 \times 1000$. The results at $t=0.8$ are presented. Fig. \ref{2r_density} shows the density distribution. Fig. \ref{2r_vorticity} shows the vorticity magnitude distribution. Good agreements are reached with the results in Ref. \cite{benchmark}. The small vortices induced by flow instabilities are well resolved on the $1000 \times 1000$ mesh.

\subsection{Laminar boundary layer}

\begin{figure}[htbp]
\centering
\begin{minipage}[t]{0.5\textwidth}
\flushright
\includegraphics[width=\textwidth]{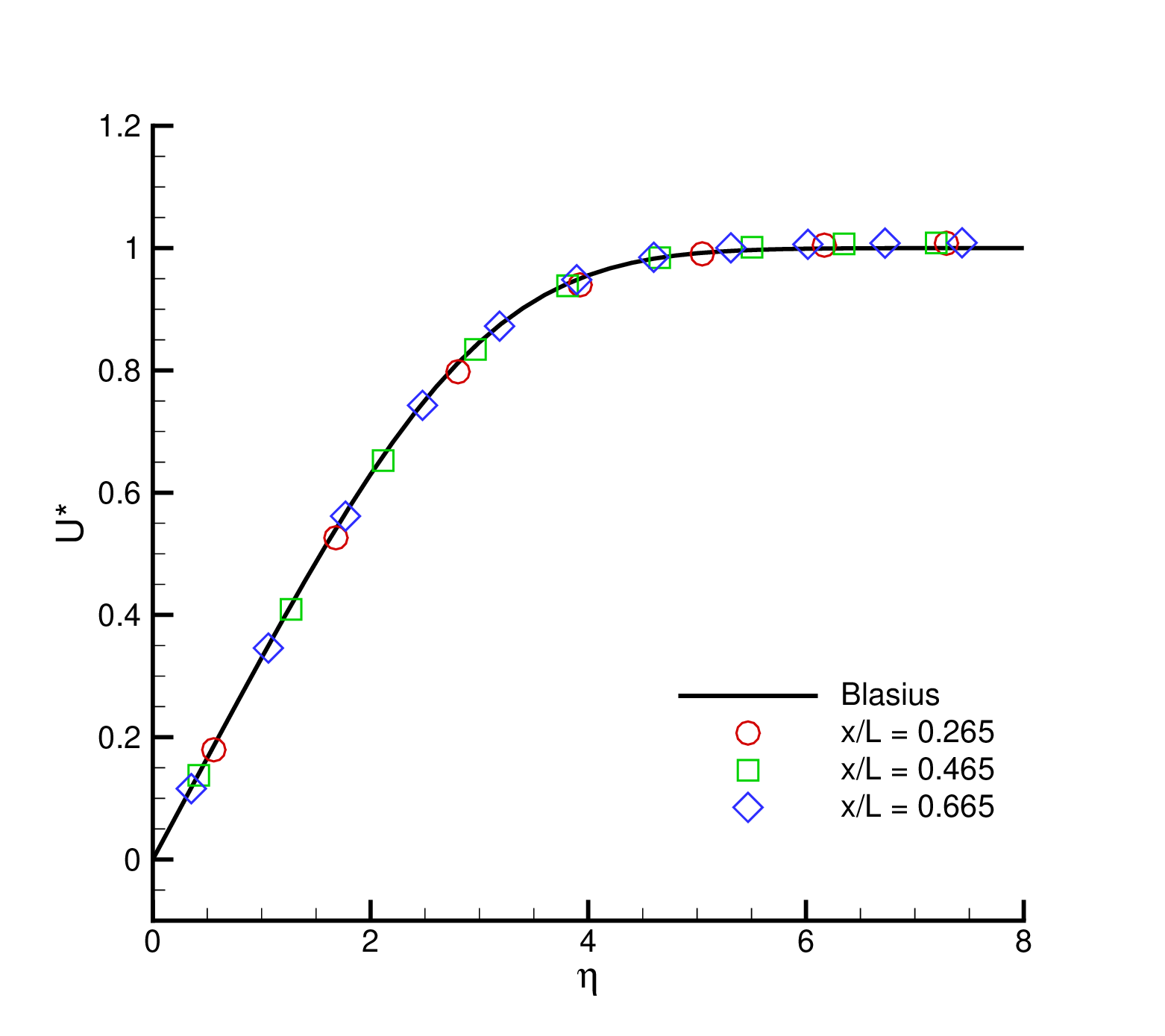}
\end{minipage}%
\begin{minipage}[t]{0.5\textwidth}
\flushleft
\includegraphics[width=\textwidth]{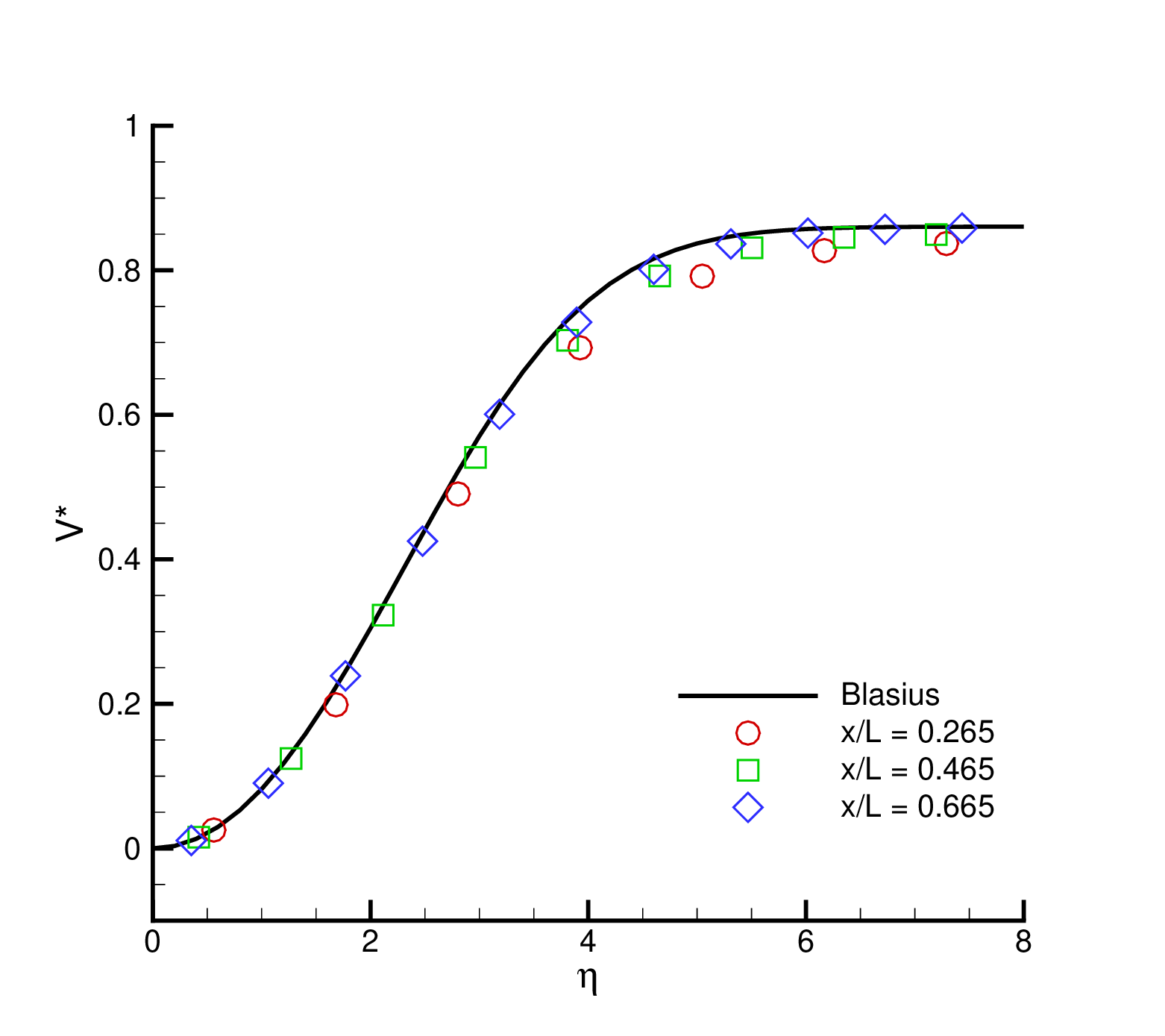}
\end{minipage}
\caption{Non-dimensional velocitiy profiles of a laminar boundary layer.}\label{bl}
\end{figure}

The computational domain is $[-0.2, 1]\times[0, 0.5]$, a flat plate with length $L=1$ is located from $x=0$. $360 \times 150$ mesh cells are uniformly distributed with $60 \times 150$ cells ahead of the plate. The mean flow Mach number is $0.15$, the Reynolds number is $Re = U_\infty L / \nu =3 \times 10^4$. At the lower boundary, the symmetric Euler reflection boundary condition is adopted for $x < 0$ and the non-slip condition is used for $x \geq 0$. At the right boundary, the simple extrapolation is used. At other boundaries the non-reflection boundary condition based on Riemann invariants is applied. Fig. \ref{bl} shows the velocity profiles at three different locations. The non-dimensional variables are defined as $U^\ast = U/U_\infty$, $V^\ast = V/\sqrt{\nu U_\infty / x}$ and $\eta = y / \sqrt{\nu x / U_\infty}$. The results of the present method have good agreements with the analytical Blasius solution. The velocity profile can be resolved with as few as $5$ grid points.

\subsection{Low speed lid-driven cavity flow}
\begin{figure}[htbp]
\centering
\begin{minipage}[t]{0.5\textwidth}
\flushright
\includegraphics[width=\textwidth]{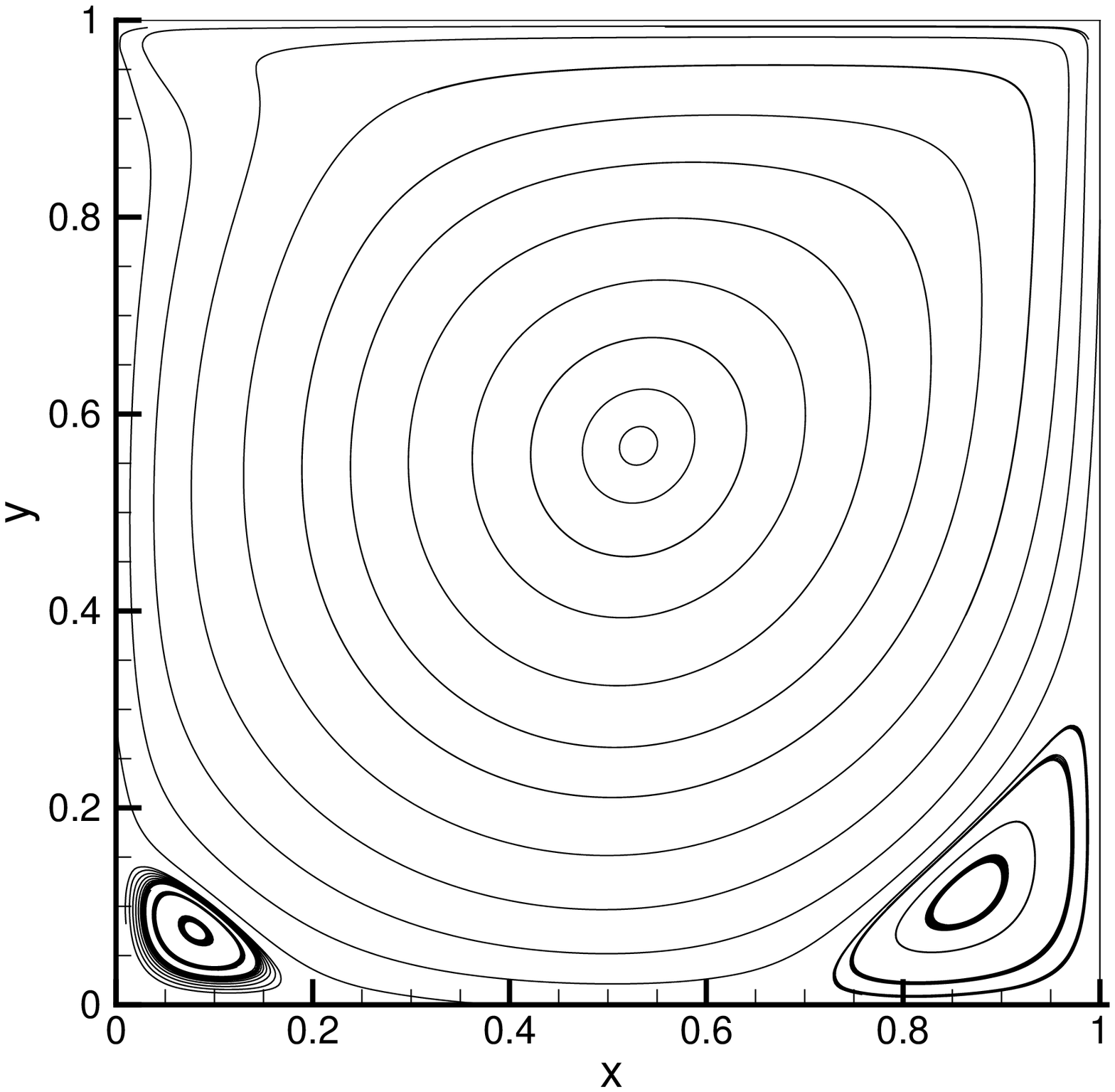}
\end{minipage}%
\begin{minipage}[t]{0.5\textwidth}
\flushleft
\includegraphics[width=\textwidth]{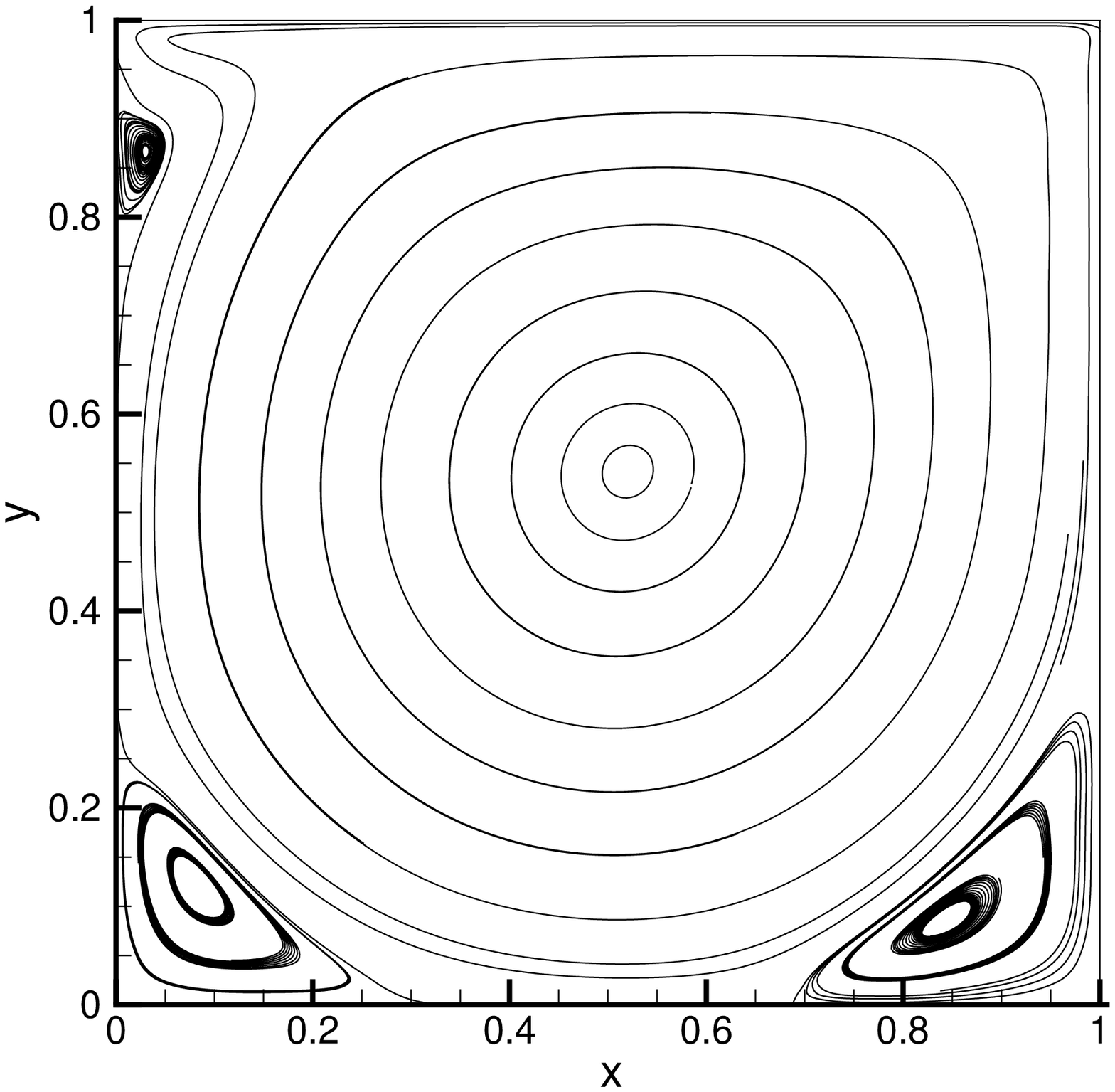}
\end{minipage}
\caption{Streamlines of the lid-driven cavity flow with $Re = 1000$ (left) and $Re = 3200$ (right).}\label{cavity_streamline}
\end{figure}

\begin{figure}[htbp]
\centering
\begin{minipage}[t]{0.5\textwidth}
\flushright
\includegraphics[width=\textwidth]{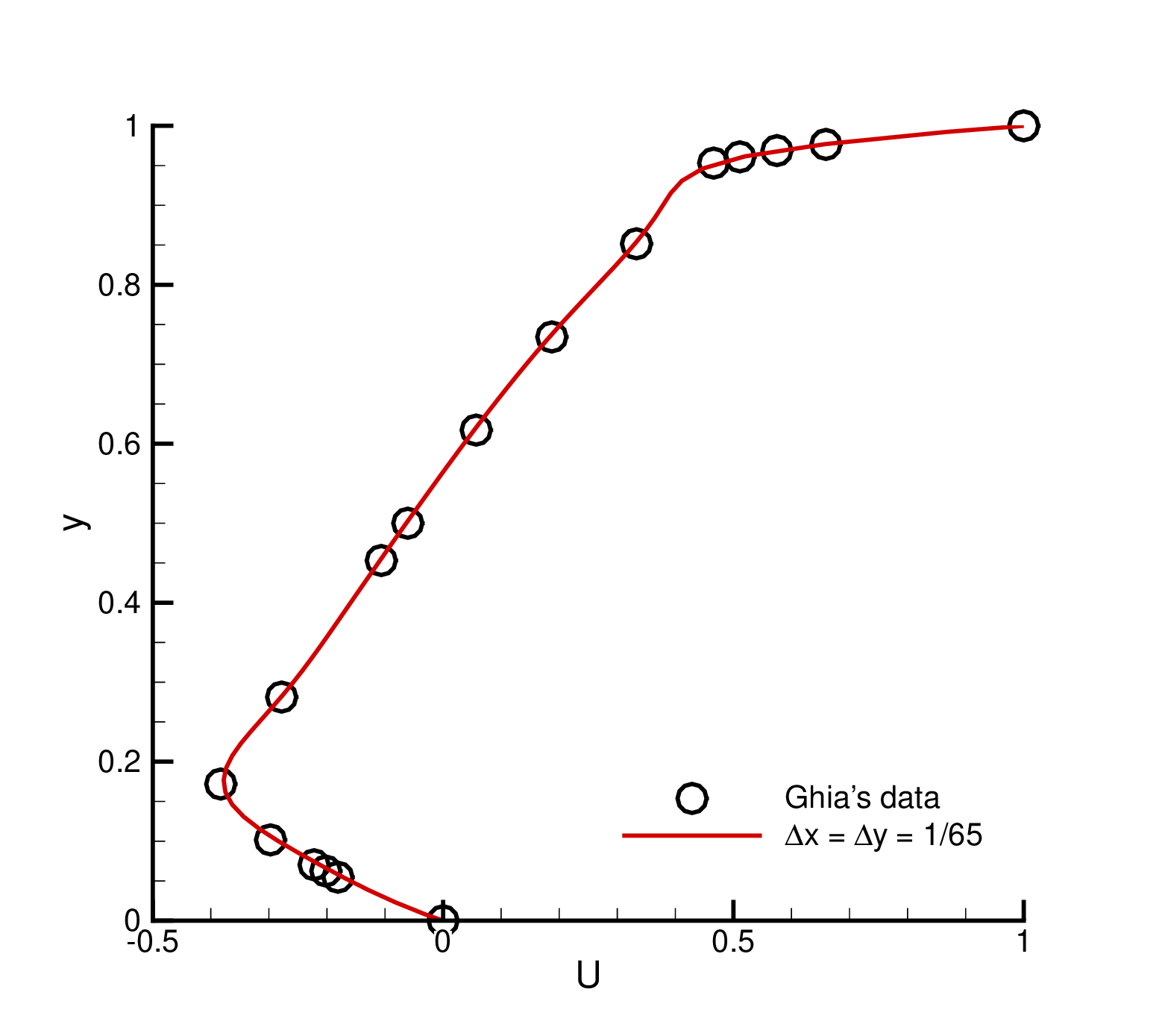}
\end{minipage}%
\begin{minipage}[t]{0.5\textwidth}
\flushleft
\includegraphics[width=\textwidth]{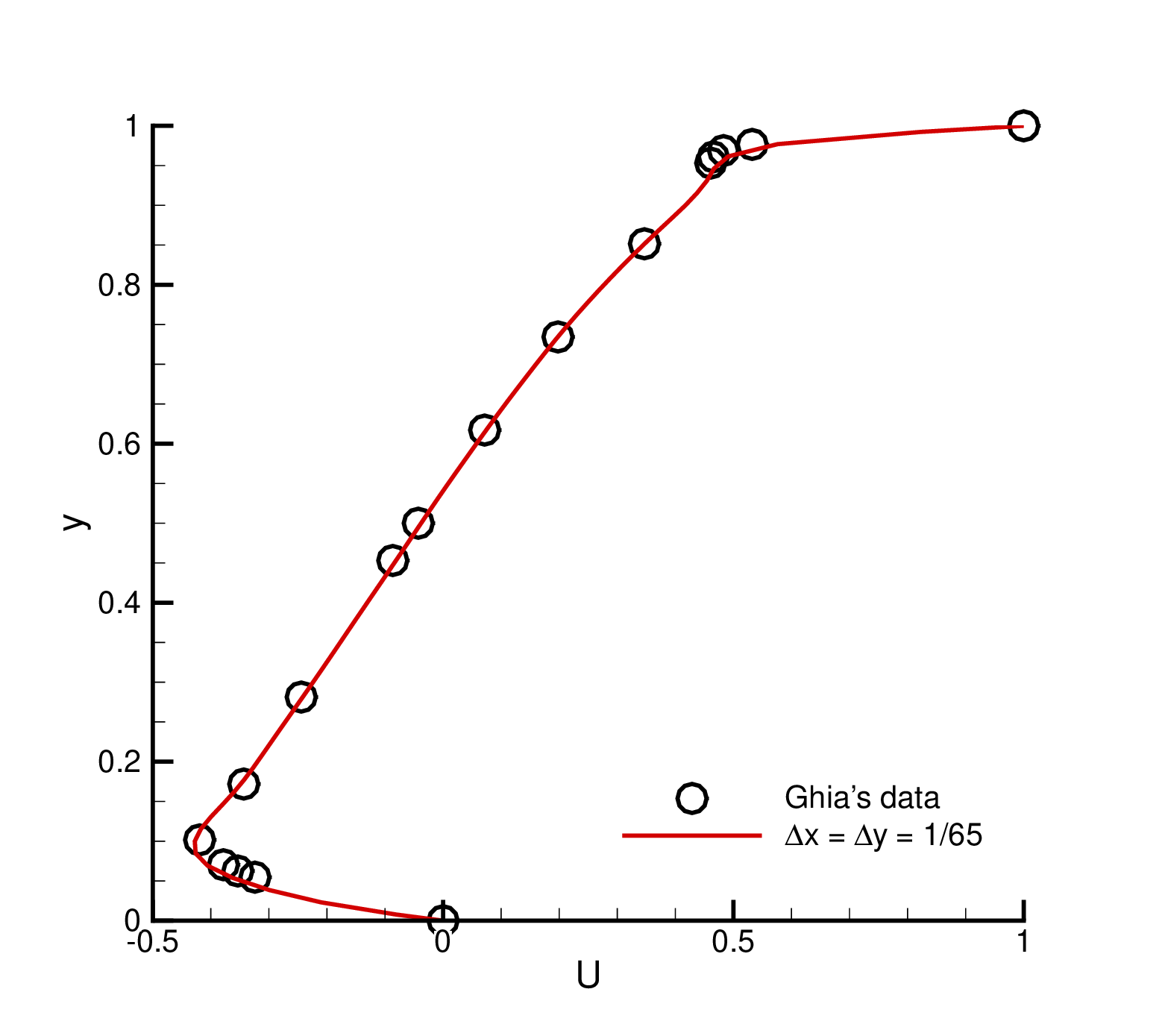}
\end{minipage}
\begin{minipage}[t]{0.5\textwidth}
\flushright
\includegraphics[width=\textwidth]{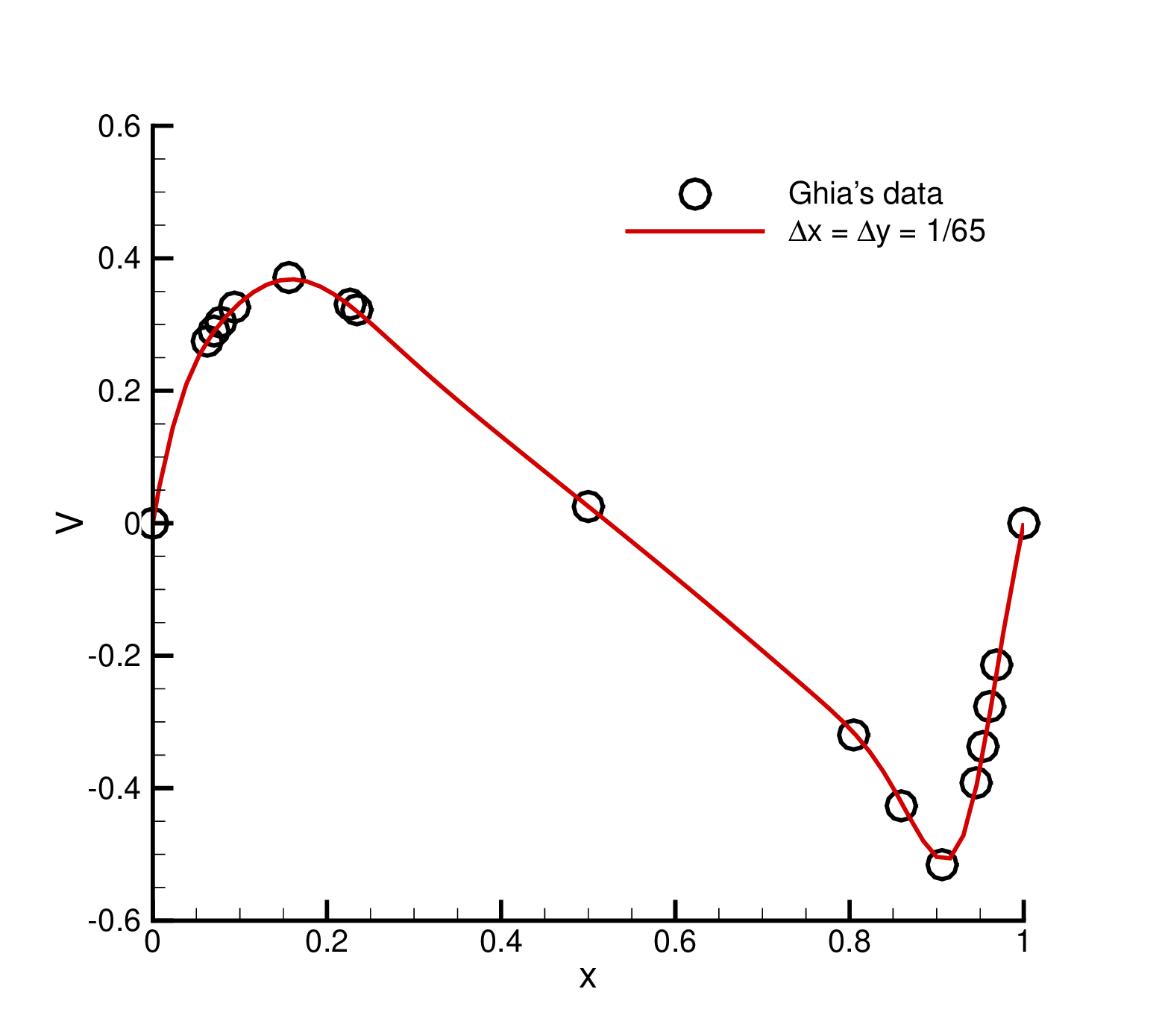}
\end{minipage}%
\begin{minipage}[t]{0.5\textwidth}
\flushleft
\includegraphics[width=\textwidth]{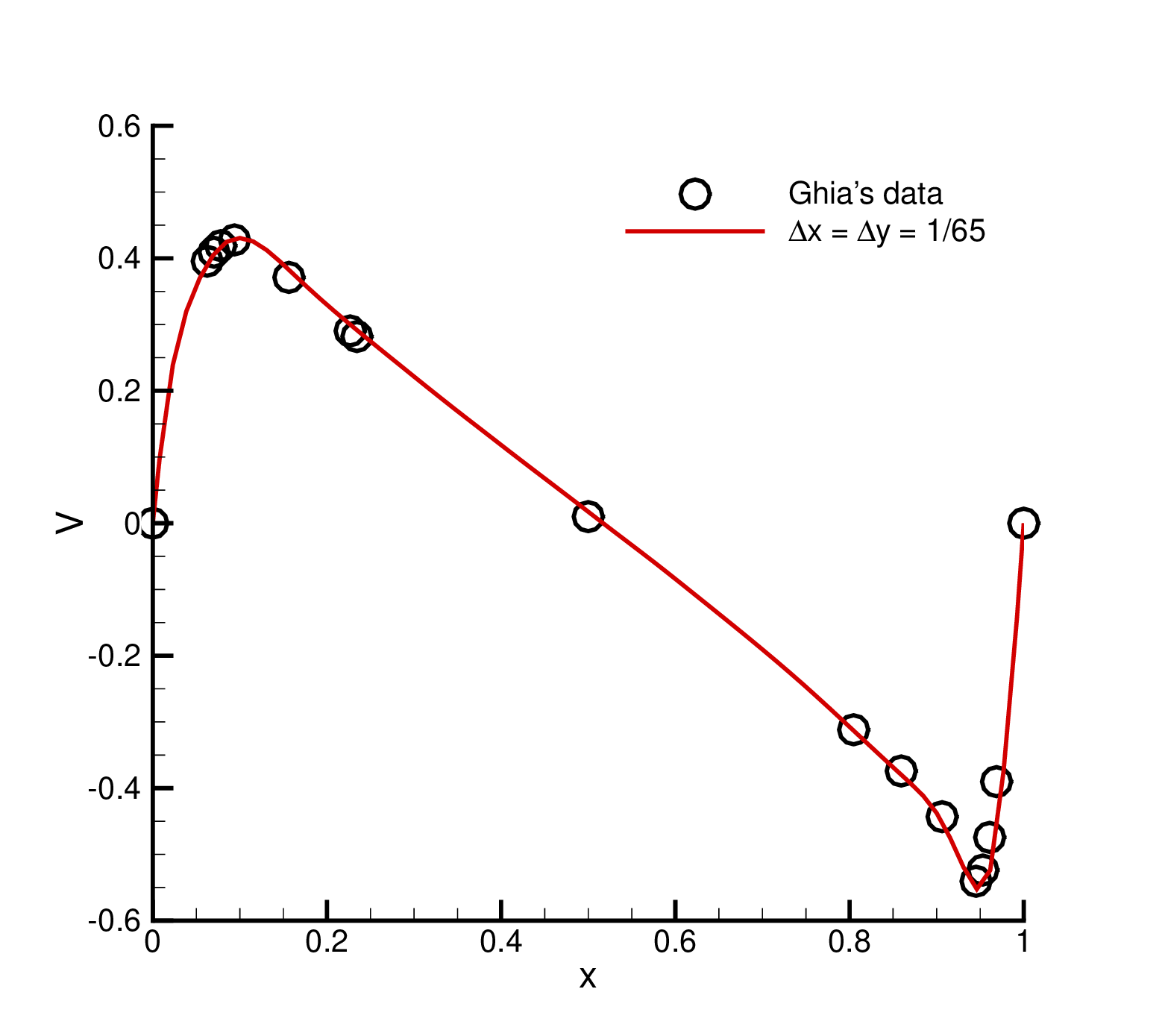}
\end{minipage}
\caption{Velocity profile of the cavity flow with $Re = 1000$ (left) and $Re = 3200$ (right). $U$ velocity is taken at the line $x=0.5$, and $V$ velocity is taken at the line $y=0.5$.}\label{cavity_uv}
\end{figure}

The fluid is bounded in a cavity with unit side length and driven by the top lid. The lid moves to the right with a uniform velocity. The Mach number is set to be $0.3$, the adiabatic non-slip condition is applied for all boundaries. Since it is a test case for incompressible flow, most simulations in the past use continuous initial reconstruction at the cell interface to minimize the kinematic dissipation \cite{Xu2003}. However, here we still use the shock capturing WENO reconstruction, which leads to a discontinuous initial state at the interface. This case is run at $Re=1000$ and $Re = 3200$. For both cases we use a $65 \times 65$ mesh. The reference data is from Ref. \cite{Ghia1982}. The streamlines are presented in Fig. \ref{cavity_streamline}, where the vortex structures are very clear. Fig. \ref{cavity_uv} shows the velocity profiles at the central lines of the cavity. For a Reynolds number as high as $3200$, the results of a $65 \times 65$ mesh still match the reference data very well.

\subsection{The viscous shock tube problem}

We consider the test case studied by Daru and Tenaud \cite{Daru2001}. A diaphragm is vertically located in the middle of a square 2-D shock tube with unit side length, separating the space into the left and right parts.  The initial state in the non-dimensional form is given by 
\begin{equation}
\left( \rho, U, V, p \right) = \left\{ \begin{aligned}
& \left( 120, 0, 0, 120/\gamma \right), \quad 0 \leq x \leq 0.5, \\
& \left( 1.2, 0, 0, 1.2/\gamma \right), \quad  0.5 < x \leq 1.
\end{aligned} \right.
\end{equation}

For air, $\gamma = 1.4$ and the Prandtl number $Pr = 0.73$. All boundaries of the tube is non-slip and adiabatic. When the diaphragm is broken at $t = 0$, a shock forms and then moves towards right, followed by a contact discontinuity. The Mach number of the shock is $2.37$. Simultaneously, a rarefaction wave propagates towards left. After reaching the right wall, the shock is reflected back and moves to the left, interacting with the contact discontinuity and the rarefaction wave. In viscous cases, a thin boundary layer is generated by the shear between the shock and the horizontal boundaries, resulting in complex shock wave/boundary layer interactions. Detailed analyses of the process can be found in Ref. \cite{Daru2009}.

\begin{figure}[htbp]
\centering
\includegraphics[width=0.8\textwidth]{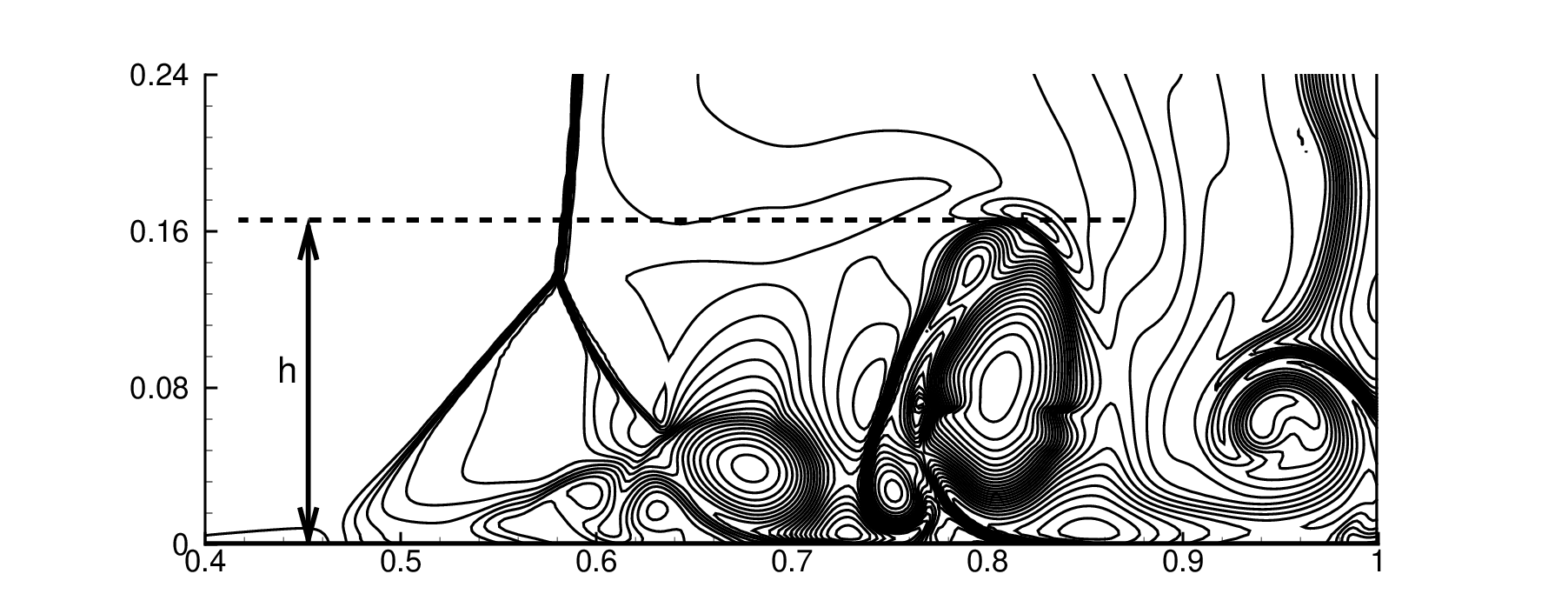}
\caption{Density distribution at $t=1$ of the viscous shock tube problem at $Re= 200$. The mesh used is $500 \times 250$.}\label{vst200}
\end{figure}

The non-dimensional time for comparison is $t = 1$. Only the domain $[0,1] \times [0,0.5]$ is computed due to symmetry of the problem. The CFL number in this case is $1$. First we consider the case with $Re=200$. A $500 \times 250$ mesh is employed. The density distribution is shown in Fig. \ref{vst200}. The height of the primary vortex is compared with that by reference methods in Ref. \cite{Kim2005, Pan2016}, as listed in Table \ref{table height}. A good agreement is reached.

\begin{table}[htbp]
\centering
\begin{tabular}{ccccc}
\hline
Scheme        & AUSMPW+ \cite{Kim2005} & M-AUSMPW+ \cite{Kim2005}  &   fourth-order GKS \cite{Pan2016}  & present \\
\hline
Height ($h$) & $0.163$         & $0.166$              & $0.171$    &   $0.166$  \\
\hline
\end{tabular}
\caption{Comparison of the primary vortex height at $Re=200$ for different schemes on a $500 \times 250$ mesh.} \label{table height}
\end{table}

\begin{figure}[htbp]
\centering
\includegraphics[width=0.8\textwidth]{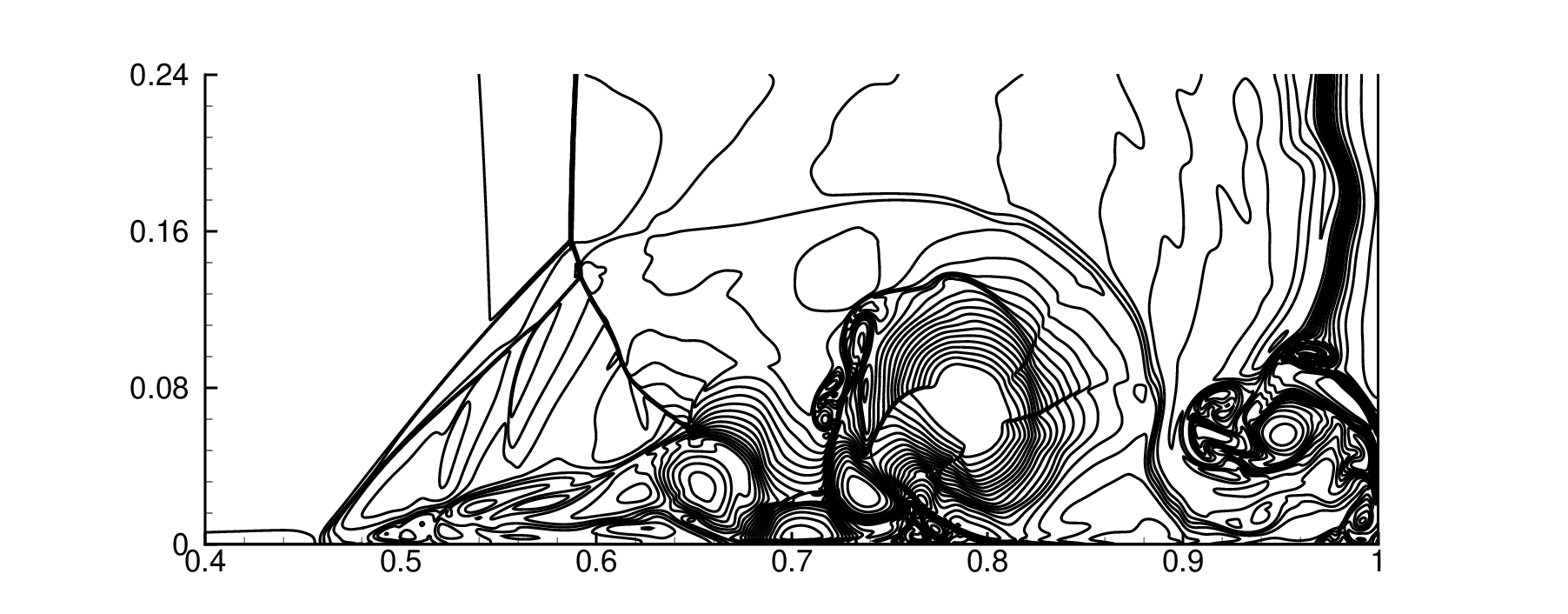}
\caption{Density distribution at $t=1$ of the viscous shock tube problem at $Re= 1000$. The mesh used is $2000 \times 1000$.}\label{vst1000}
\end{figure}

The flow becomes much more complex at a higher Reynolds number $Re = 1000$. Fig. \ref{vst1000} shows the density distribution at $t=1$ with a $2000 \times 1000$ mesh. The small-scale structures are clearly presented. The distribution is very much similar to the result in Ref. \cite{Daru2009}.

\section{Comparison of Accuracy and Efficiency} \label{section efficiency}
To compare the accuracy and efficiency of the simplified method with the baseline method (WENO-GKS in Ref. \cite{Luo2013a}), two test cases are selected from Section \ref{section cases}. We ensure that all conditions except the evolution model are exactly the same for all computations. And all cases in this section are run on the same laptop. Three methods are compared in this section:
\begin{itemize}
\item The baseline method. This is the original method in Ref. \cite{Luo2013a}. It uses the distribution function in Eq. \eqref{before_simplification} and the coefficients in Eq. \eqref{coeff}.
\item The S1 method. This method is the one with Simplification 1 but without Simplification 2. It uses the distribution function in Eq. \eqref{before_simplification} and the coefficients in Eq. \eqref{coeff_new}.
\item The present method. This is the final simplified scheme. It uses the distribution function in Eq. \eqref{simplified} and the coefficients in Eq. \eqref{coeff_new}.
\end{itemize}

\subsection{1-D inviscid case: the Shu-Osher problem} 
\begin{figure}[htbp]
\centering
\begin{minipage}[t]{0.5\textwidth}
\flushright
\includegraphics[width=\textwidth]{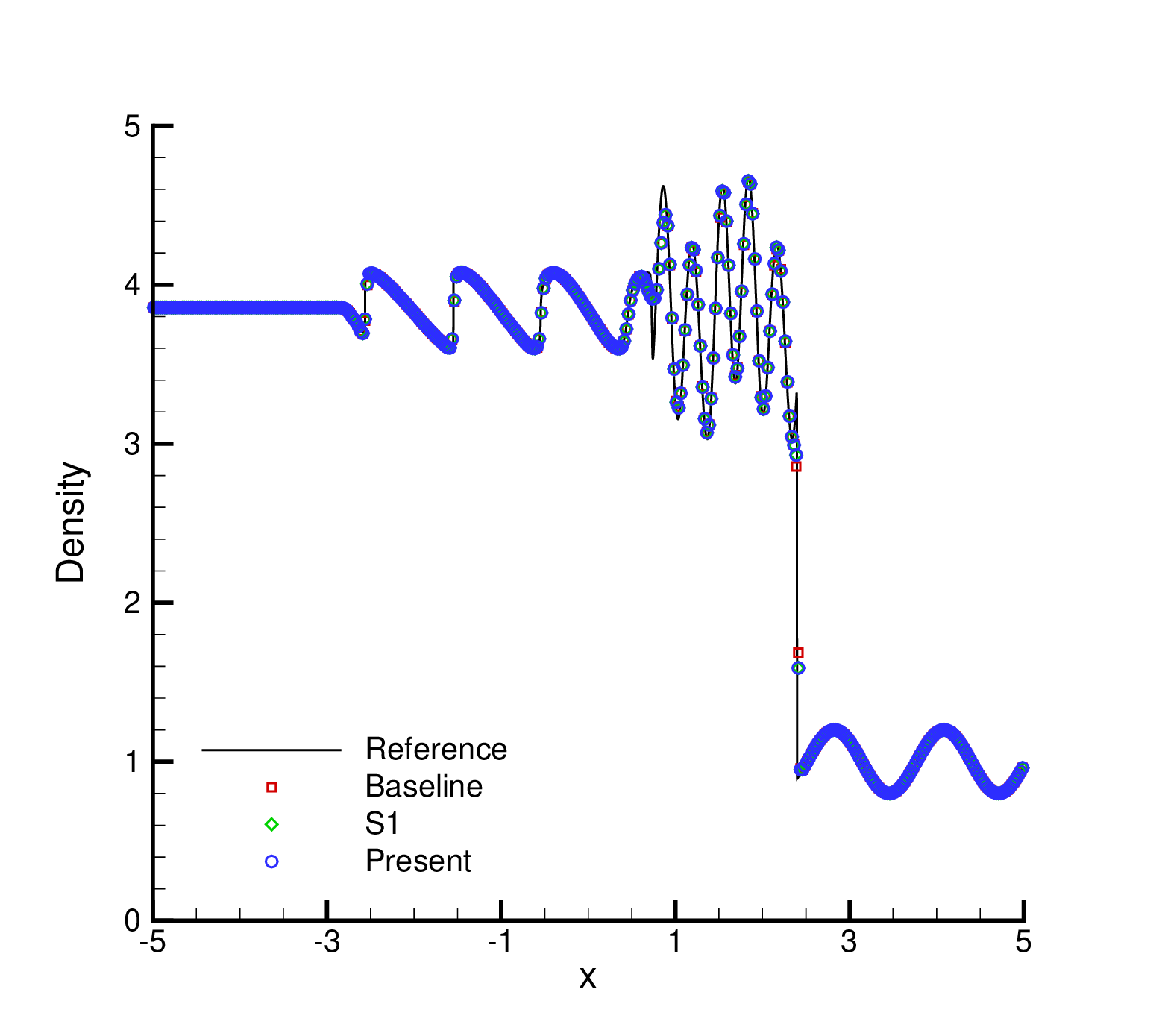}
\end{minipage}%
\begin{minipage}[t]{0.5\textwidth}
\flushleft
\includegraphics[width=\textwidth]{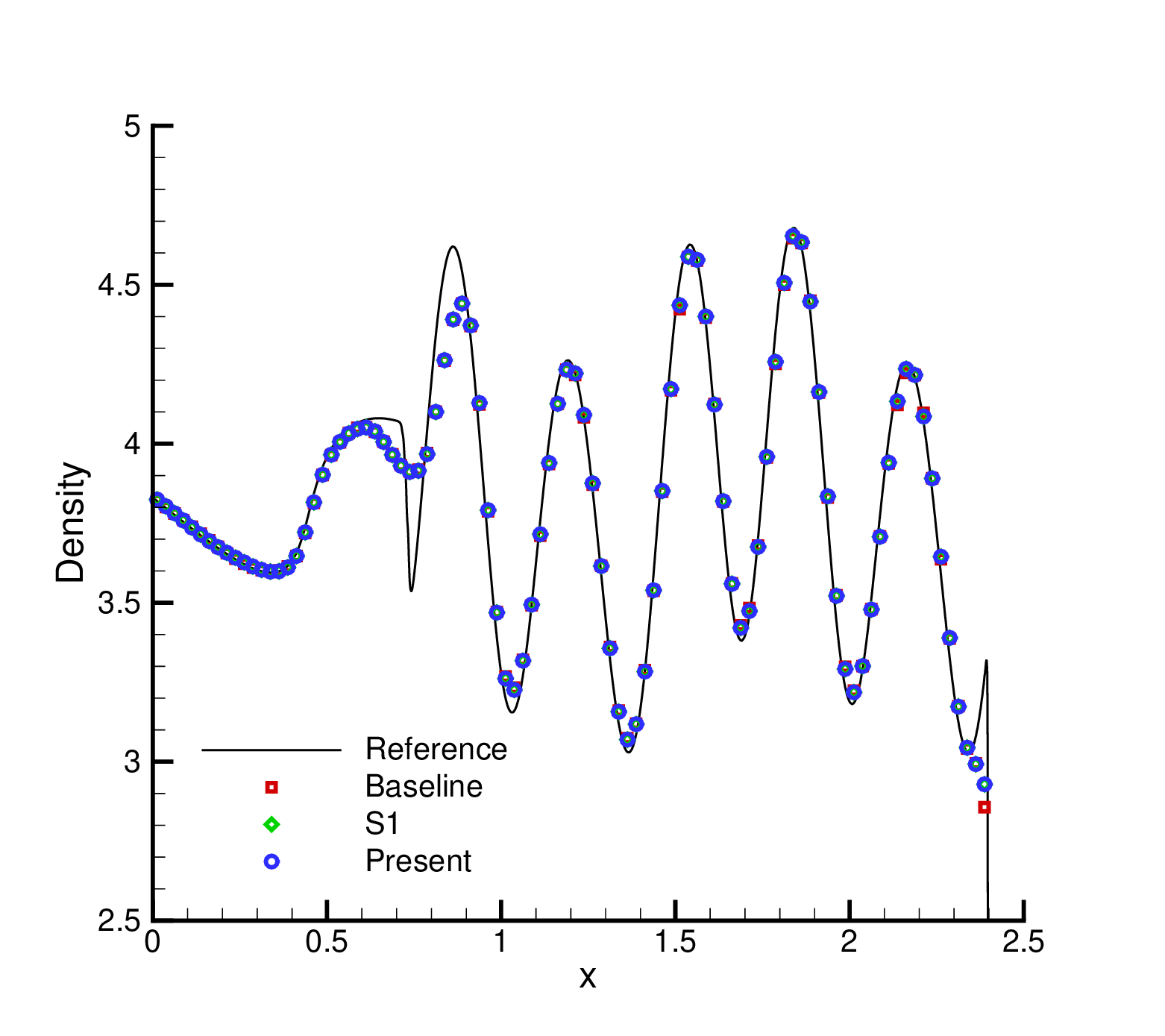}
\end{minipage}
\caption{Density distribution at $t = 1.8$ of the Shu-Osher problem. $400$ cells are used.}\label{shu_compare}
\end{figure}

The density distributions at $t=1.8$ of the Shu-Osher problem by the three methods are compared in Fig. \ref{shu_compare}. Notice that this is an inviscid case, and no artificial dissipation is added during the computation. So Simplification 2 has no effect since it is a modification related to the viscous terms. Then the results of the S1 method and the present method are exactly the same. The results of the baseline method and the present method are also very similar. Both can resolve the high-frequency oscillations very well. Small differences are visible only near the strong discontinuity around $x=2.4$.

The computational efficiency is shown in Table \ref{table shu_efficiency}. The CPU time for a complete computation (to the non-dimensional time $t=1.8$) with $1000$ cells is compared. For one-dimensional cases, the speedup after the simplification is $267\%$.

\begin{table}[htbp]
\centering
\begin{tabular}{c|c|c}
\hline
Method & CPU time (seconds) & Speedup   \\
\hline
Baseline & $13.67$ & $-$\\
\hline
S1 & $5.34$  & $156\%$ \\
\hline
Present & $3.72$ & $267\%$ \\
\hline
\end{tabular}
\caption{Comparison of efficiency for different methods based on the Shu-Osher problem.} \label{table shu_efficiency}
\end{table}

\subsection{2-D viscous case: the viscous shock tube problem} 
The viscous shock tube problem is considered for the two-dimensional viscous cases. We choose this case because it is very sensitive to the choice of computational methods. 

\begin{figure}[htbp]
\centering
\begin{minipage}[t]{0.5\textwidth}
\flushright
\includegraphics[width=\textwidth]{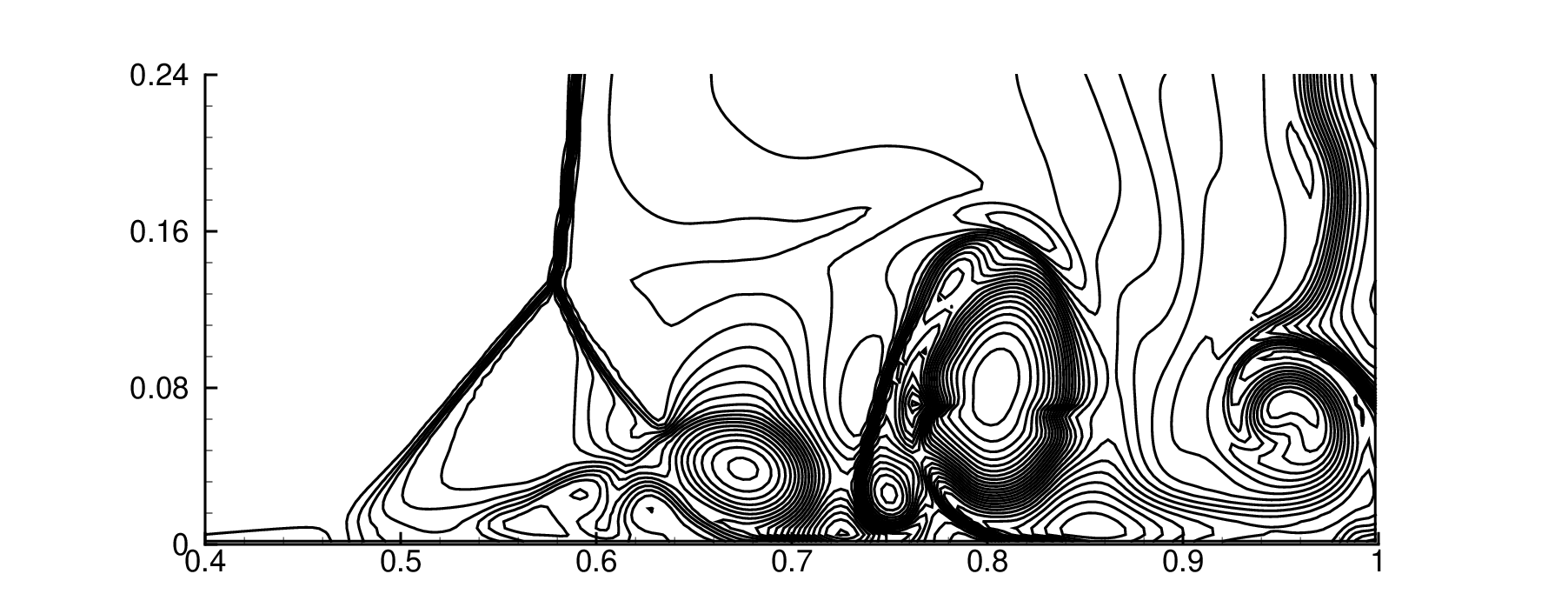}
\end{minipage}%
\begin{minipage}[t]{0.5\textwidth}
\flushleft
\includegraphics[width=\textwidth]{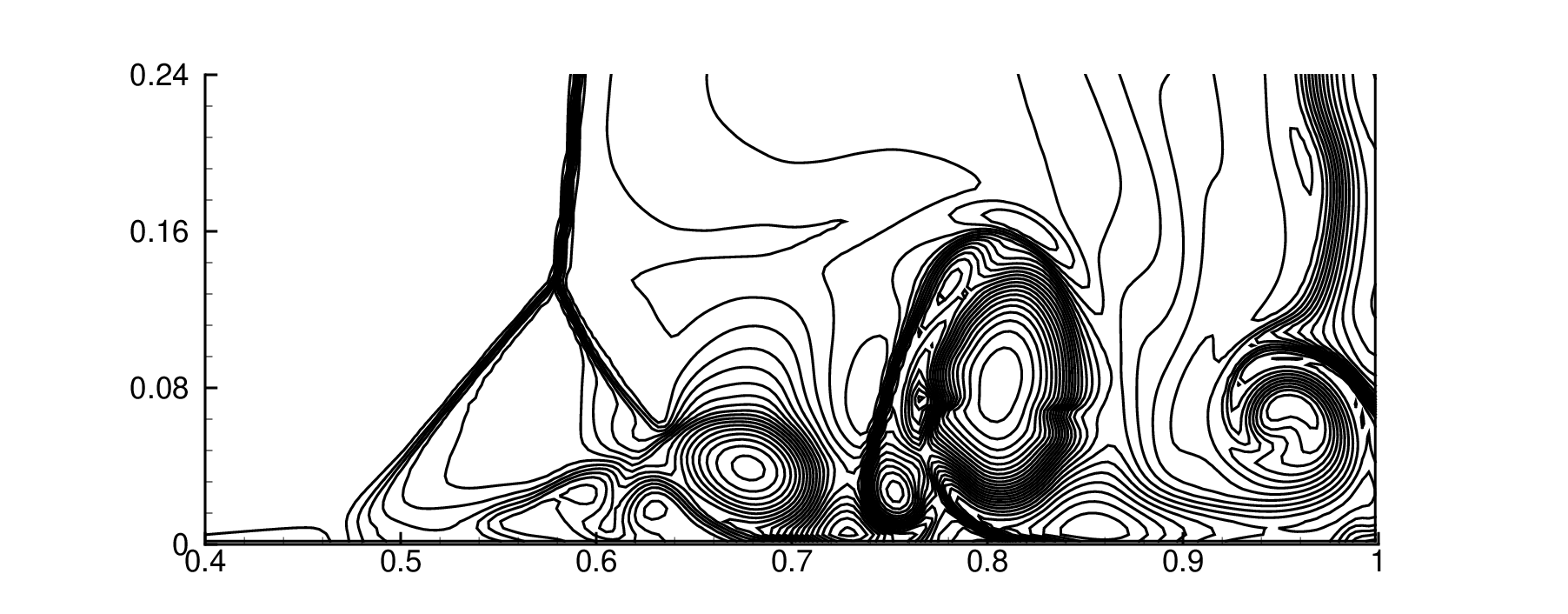}
\end{minipage}
\caption{Density distribution at $t=1$ of the viscous shock tube problem at $Re=200$ with a $300 \times 150$ mesh. Left: the baseline method. Right: the present method.}\label{vst_compare}
\end{figure}

The $Re=200$ case is computed with the baseline method and the present method on a $300 \times 150$ mesh. The results are in Fig. \ref{vst_compare}. The difference is almost indistinguishable. 

\begin{figure}[htbp]
\centering
\includegraphics[width=0.7\textwidth]{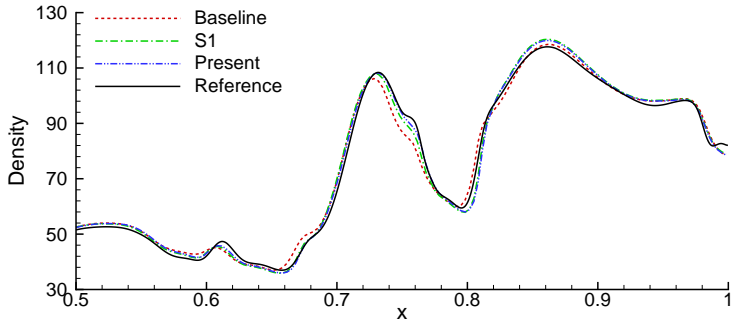}
\caption{Density distribution along the bottom wall at $t=1$ of the viscous shock tube problem. $Re= 200$. The mesh used is $300 \times 150$.}\label{vst_wall_compare_mesh300}
\end{figure}

\begin{figure}[htbp]
\centering
\includegraphics[width=0.7\textwidth]{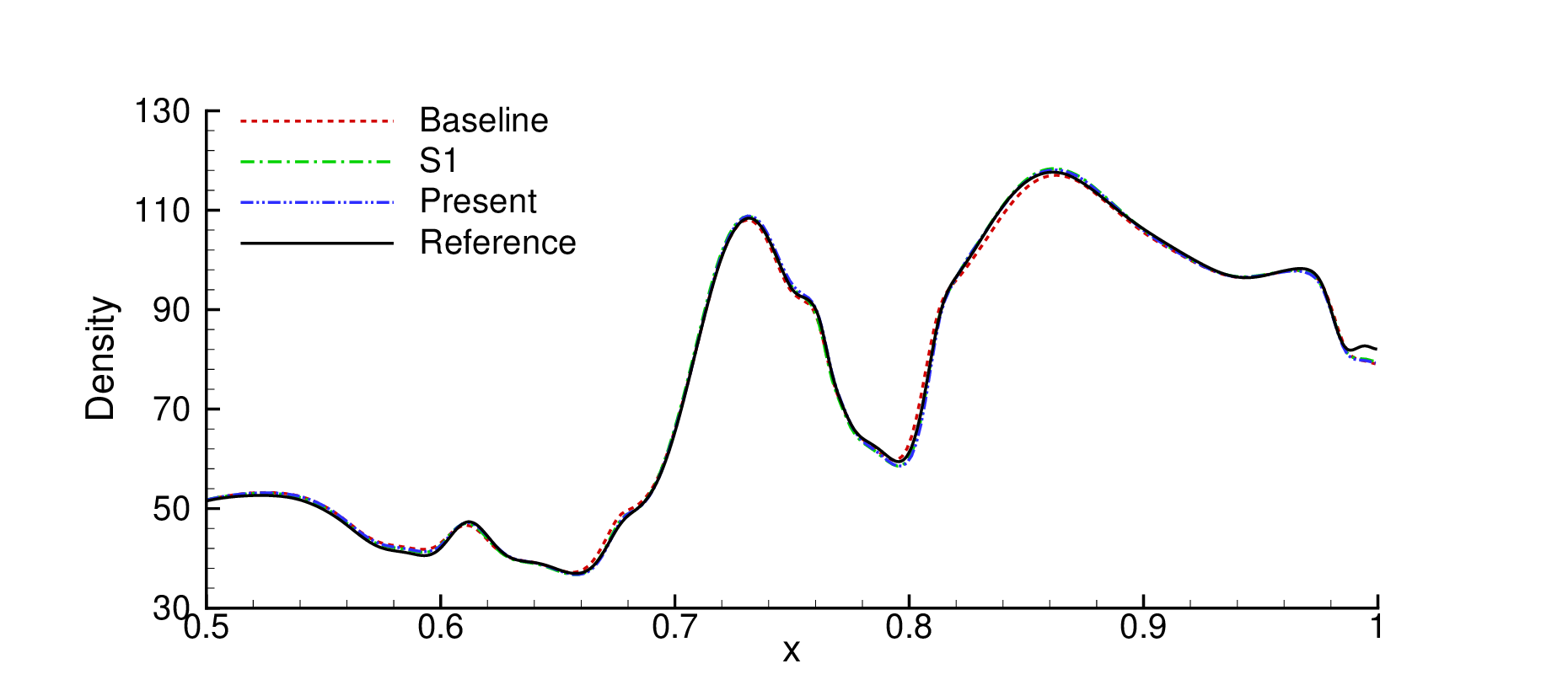}
\caption{Density distribution along the bottom wall at $t=1$ of the viscous shock tube problem. $Re= 200$. The mesh used is $500 \times 250$.}\label{vst_wall_compare_mesh500}
\end{figure}

For a more accurate pointwise comparison, the density distributions along the bottom wall at $t=1$ are plotted in Fig. \ref{vst_wall_compare_mesh300} (on the $300 \times 150$ mesh) and Fig. \ref{vst_wall_compare_mesh500} (on a $500 \times 250$ mesh). The reference curve is a converged solution on a very fine ($1500 \times 750$) mesh. It is found that some differences exist on the coarse mesh. But when the mesh is refined, the difference becomes very small. On both meshes, the results of the S1 method and the present method are very similar, and the present method is no worse than the baseline method. In fact, the present one is even slightly better. This phenomenon may be due to the fact that by applying Simplification 1, the computations of very high moments of the gas distribution function are avoided, so that less error is introduced during the computation (see Section \ref{section modify}). The original baseline method with many higher-order moment terms may have the dynamical effect corresponding to  the Burnett equations, which may not be physically valid \cite{Zhong1993, Xu2002}.

Table \ref{vst_efficiency} shows the efficiency comparison. The CPU time for one time step on a $1000 \times 500$ mesh is compared. The simplified method is about $4$ times faster than the baseline method.
\begin{table}[htbp]
\centering
\begin{tabular}{c|c|c}
\hline
Method & CPU time (seconds) & Speedup   \\
\hline
Baseline & $18.62$ & $-$\\
\hline
S1&   $6.68$  & $179\%$ \\
\hline
Present & $3.76$ & $395\%$ \\
\hline
\end{tabular}
\caption{Comparison of efficiency for different methods based on the viscous shock tube problem.} \label{vst_efficiency}
\end{table}

\subsection{Discussion}
The above numerical examples show that the efficiency improvement in two-dimensional case is larger than that in one-dimensional case. This is because for the gas-kinetic method, a 2-D scheme is not a simply doubled 1-D scheme. Derivatives in the second dimension also participate in the evolution model due to its multidimensionality, and the cross terms of space- and time-derivatives are also introduced. Similarly, when three spatial dimensions are considered, more terms will appear. Therefore, by using the simplifications in this paper, more gain in computational efficiency is expected for 3-D problems.

\section{Conclusion} \label{section conclusion}
In this paper, two simplifications have been introduced on the gas evolution model of a baseline higher-order gas-kinetic scheme with WENO reconstruction. The modifications are designed based on the physical analysis of higher-order terms in the time evolving gas distribution function and  the accuracy is not affected. The resultant method significantly reduces the computational cost. It is about $4$ times more efficient than the baseline method for two-dimensional cases. Various standard cases are tested to show the robustness, accuracy, and efficiency of the simplified scheme. The simplifications and the design of the artificial dissipation introduced in this paper can also be adopted by other kinds of high-order gas-kinetic methods.

\section*{Acknowledgements}
The current work of K. Xu is supported by Hong Kong Research Grant Council (620813, 16211014, 16207715), and National Science Foundation of China (91330203,91530319).

\bibliography{mybibfile}

\begin{thebibliography}{10}
\expandafter\ifx\csname url\endcsname\relax
  \def\url#1{\texttt{#1}}\fi
\expandafter\ifx\csname urlprefix\endcsname\relax\def\urlprefix{URL }\fi
\expandafter\ifx\csname href\endcsname\relax
  \def\href#1#2{#2} \def\path#1{#1}\fi

\bibitem{Xu1998}
K.~Xu, Gas-kinetic schemes for unsteady compressible flow simulations, von
  Karman Institute report (1998).

\bibitem{Xu2001}
K.~Xu, A gas-kinetic \text{BGK} scheme for the \text{Navier-Stokes} equations
  and its connection with artificial dissipation and \text{Godunov} method,
  Journal of Computational Physics 171 (2001) 289--335.

\bibitem{Liu1994}
X.-D. Liu, S.~Osher, T.~Chan, Weighted essentially non-oscillatory schemes,
  Journal of Computational Physics 115 (1994) 200--212.

\bibitem{Jiang1996}
G.-S. Jiang, C.-W. Shu, Efficient implementation of weighted \text{ENO}
  schemes, Journal of Computational Physics 126 (1996) 202--228.

\bibitem{Titarev2004}
V.~A. Titarev, E.~F. Toro, Finite-volume \text{WENO} schemes for
  three-dimensional conservation laws, Journay of Computational Physics 201
  (2004) 238--260.

\bibitem{Li2010}
Q.~Li, K.~Xu, S.~Fu, A high-order gas-kinetic \text{Navier-Stokes} flow solver,
  Journal of Computational Physics 229 (2010) 6715--6731.

\bibitem{Luo2013a}
J.~Luo, K.~Xu, A high-order multidimensional gas-kinetic scheme for
  hydrodynamic equations, Science China Technological Sciences 56 (2013)
  2370--2384.

\bibitem{Liu2014}
N.~Liu, H.~Tang, A high-order accurate gas-kinetic scheme for one- and
  two-dimensional flow simulation, Communications in Computational Physics 15
  (2014) 911--943.

\bibitem{Luo2013b}
J.~Luo, L.~Xuan, K.~Xu, Comparison of fifth-order \text{WENO} scheme and finite
  volume \text{WENO}-gas-kinetic scheme for inviscid and viscous flow
  simulation, Communications in Computational Physics 14 (2013) 599--620.

\bibitem{BGK}
P.~L. Bhatnagar, E.~P. Gross, M.~Krook, A model for collision processes in
  gases. \text{I}. \text{Small} amplitude processes in charged and neutral
  one-component systems, Physical Review 94 (1954) 511--525.

\bibitem{Shu1997}
C.-W. Shu, Essentially non-oscillatory and weighted essentially non-oscillatory
  schemes for hyperbolic conservation laws, Institute for Computer Applications
  in Science and Engineering (1997).

\bibitem{Ohwada2004}
T.~Ohwada, K.~Xu, The kinetic scheme for the full-\text{Burnett} equations,
  Journal of Computational Physics 201 (2004) 315--332.

\bibitem{Pan2016}
L.~Pan, K.~Xu, Q.~Li, J.~Li, An efficient and accurate two-stage fourth-order
  gas-kinetic scheme for the \text{Euler} and \text{Navier-Stokes} equations,
  Journal of Computational Physics 326 (2016) 197--221.

\bibitem{Woodward1984}
P.~Woodward, P.~Colella, The numerical simulation of two-dimensional fluid flow
  with strong shocks, Journal of Computational Physics 54 (1984) 115--173.

\bibitem{Shu1989}
C.-W. Shu, S.~Osher, Efficient implementation of essentially non-oscillatory
  shock-capturing schemes \text{II}, Journal of Computational Physics 83 (1989)
  32--78.

\bibitem{Zhang1990}
T.~Zhang, Y.~Zheng, Conjecture on the structure of solutions of the
  \text{Riemann} problem for two-dimensional gas dynamics systems, SIAM Journal
  on Mathematical Analysis 21 (1990) 593--630.

\bibitem{benchmark}
L.~Pan, J.~Li, K.~Xu, A few benchmark test cases for higher-order \text{Euler}
  solvers, arXiv:1609.04491v1 [math.NA], 15 Sept. 2016.

\bibitem{Xu2003}
K.~Xu, X.~He, Lattice \text{Boltzmann} method and gas-kinetic \text{BGK} scheme
  in the low-\text{Mach} number viscous flow simulations, Journal of
  Computational Physics 190 (2003) 100--117.

\bibitem{Ghia1982}
U.~Ghia, K.~N. Ghia, C.~T. Shin, High-\text{Re} solutions for incompressible
  flow using the \text{Navier-Stokes} equations and a multigrid method, Journal
  of Computational Physics 48 (1982) 387--411.

\bibitem{Daru2001}
V.~Daru, C.~Tenaud, Evaluation of \text{TVD} high resolution schemes for
  unsteady viscous shocked flows, Computers \& Fluids 30 (2001) 89--113.

\bibitem{Daru2009}
V.~Daru, C.~Tenaud, Numerical simulation of the viscous shock tube problem by
  using a high resolution monotonicity-preserving scheme, Computers \& Fluids
  38 (2009) 664--676.

\bibitem{Kim2005}
K.~H. Kim, C.~Kim, Accurate, efficient and monotonic numerical methods for
  multi-dimensional compressible flows: \text{Part I}: \text{Spatial}
  discretization, Journal of Computational Physics 208 (2005) 527--569.

\bibitem{Zhong1993}
X.~Zhong, R.~W. MacCormack, D.~R. Chapman, Stabilization of the \text{Burnett}
  equations and application to hypersonic flows, AIAA Journal 31 (1993)
  1036--1043.

\bibitem{Xu2002}
K.~Xu, Regularization of the \text{Chapman-Enskog} expansion and its
  description of shock structure, Physics of Fluids 14 (2002) 17--20.

\end{thebibliography}

\end{document}